\documentclass[letterpaper, journal, final, 10pt]{IEEEtran}

\IEEEoverridecommandlockouts
\usepackage{times}
\usepackage{amsfonts,amssymb}
\usepackage[cmex10]{amsmath}
\usepackage{array}
\usepackage{lineno}
\usepackage{graphicx,cite}
\usepackage[noend]{algorithmic}
\usepackage{algorithm}
\usepackage{verbatim}
\usepackage{bbm}
\usepackage{url}
\usepackage{subfigure}
\usepackage[normalem]{ulem}
\usepackage[usenames,dvipsnames]{color}

\newtheorem{theorem}{Theorem}[section]
\newtheorem{corollary}[theorem]{Corollary}
\newtheorem{lemma}{Lemma}
\newtheorem{proposition}[theorem]{Proposition}

\newtheorem{remark}{Remark}[section]

\def\NN{{\mathbb N}}
\def\EE{{\mathbb E}}
\def\PP{{\mathbb P}}
\def\RR{{\mathbb R}}

\def\ind{{\mathbbmss{1}}}

\def\Pcal{{\mathcal P}}

\def\kl{\mathrm{KL}}
\def\klg{\mathrm{KLG}}

\newcommand{\bI}{\mathbbmss{1}}

\newcommand{\bp}{\noindent{\it Proof.}\ }
\newcommand{\ep}{\hfill $\Box$}
\newcommand{\BEAS}{\begin{eqnarray*}}
\newcommand{\EEAS}{\end{eqnarray*}}
\newcommand{\BEA}{\begin{eqnarray}}
\newcommand{\EEA}{\end{eqnarray}}
\newcommand{\BEQ}{\begin{equation}}
\newcommand{\EEQ}{\end{equation}}
\newcommand{\BIT}{\begin{itemize}}
\newcommand{\EIT}{\end{itemize}}
\newcommand{\BNUM}{\begin{enumerate}}
\newcommand{\ENUM}{\end{enumerate}}
\newcommand{\als}[1]{ \begin{align*} #1  \end{align*}}
\newcommand{\eqs}[1]{ \begin{equation*} #1  \end{equation*}}

\newcommand{\eq}[1]{ \begin{equation} #1  \end{equation}}

\newcommand{\sk}{\nonumber\\}
\newcommand{\eqdist}{\,{\buildrel d \over =}\,}

\graphicspath{{./figures/}}

\title{Stochastic Online Shortest Path Routing: \\ The Value of Feedback}

\author{
M. Sadegh Talebi, Zhenhua Zou, Richard Combes, Alexandre Proutiere, and Mikael Johansson
\thanks{A preliminary version of this work was presented at the 2014 American Control Conference~\cite{ZouACC2014}.

M.~S.~Talebi, A.~Proutiere, and M.~Johansson are with the ACCESS Linnaeus Center and the School of Electrical Engineering, KTH Royal Institute of Technology, SE-100 44 Stockholm, Sweden (e-mail: \{mstms@kth.se, alepro@kth.se, mikaelj@kth.se\}).

Z.~Zou is with Ericsson Research, Stockholm, Sweden (e-mail:
zhenhua.zou@chalmers.se).

R.~Combes is with the Telecommunications Department, Supelec, Gif-Sur-Yvette Cedex 91192, France (e-mail: richard.combes@supelec.fr).
}

}

\begin{document}
\maketitle

%%%%%%%%%%%%%%%%%%%%%%%%%%%%%%%%%%%%%%%%%%%%%%%%%%%%%%%%%%%%%%%%%%%%%%%%%%%%%%%%
\begin{abstract}
  This paper studies online shortest path routing over multi-hop networks. Link costs or delays are time-varying and modeled by independent and identically distributed random processes, whose parameters are initially unknown. The parameters, and hence the optimal path, can only be estimated by routing packets through the network and observing the realized delays.  Our aim is to find a routing policy that minimizes the regret (the cumulative difference of expected delay) between the path chosen by the policy and the unknown optimal path. We formulate the problem as a combinatorial bandit optimization problem and consider several scenarios that differ in where routing decisions are made and in the information available when making the decisions. For each scenario, we derive a tight asymptotic lower bound on the regret that has to be satisfied by any online routing policy.  These bounds help us to understand the performance improvements we can expect when (i) taking routing decisions at each hop rather than at the source only, and (ii) observing per-link delays rather than end-to-end path delays. In particular, we show that (i) is of no use while (ii) can have a spectacular impact. Three algorithms, with a trade-off between computational complexity and performance, are proposed. The regret upper bounds of these algorithms improve over those of the existing algorithms, and they significantly outperform state-of-the-art algorithms in numerical experiments.
\end{abstract}
%%%%%%%%%%%%%%%%%%%%%%%%%%%%%%%%%%%%%%%%%%%%%%%%%%%%%%%%%%%%%%%%%%%%%%%%%%%%%%%%

\section{Introduction}

% Most real-world networks are dynamic and evolve over time. Packet losses in wireless sensor networks occur randomly and the average loss rates on links vary with time. Nodes in mobile ad-hoc networks are constantly moving, which affects the inter-node distances and thus the link parameters (e.g. the transmission success probability or average delay). The delays in overlay networks used in peer-to-peer applications change unpredictably as the load in the underlay network fluctuates. In many cases, the link parameters, i.e., the packet transmission success probabilities, are initially unknown and must be estimated by transmitting packets and observing the outcomes. When designing routing policies, we therefore need to address a challenging trade-off between exploration and exploitation:  on the one hand, it important to route packets on new or poorly known links to explore the network and ensure that the optimal path is eventually found; on the other hand, it is critical {\color{blue} that} the accumulated knowledge on link parameters is exploited so that paths with low expected delays are preferred. Of course, when the link parameters evolve over time, it becomes crucial to design algorithms that quickly learn link parameters so as to efficiently track the optimal path.

In most real-world networks, link delays vary stochastically due to unreliable links and random access protocols (e.g. in wireless networks), mobility  (e.g. in mobile ad-hoc networks), randomness of demand (e.g. in overlay networks for peer-to-peer applications), etc. In many cases, the associated parameters to links, e.g.~the packet transmission success probabilities in wireless sensor networks, are initially unknown and must be estimated by transmitting packets and observing the outcomes. When designing routing policies, we therefore need to address a challenging trade-off between exploration and exploitation:  on the one hand, it is important to route packets on new or poorly known links to explore the network and ensure that the optimal path is eventually found; on the other hand, it is critical that the accumulated knowledge on link parameters is exploited so that paths with low expected delays are preferred. When designing practical routing schemes, one is mostly concerned about the finite-time behaviour of the system and it is crucial to design algorithms that quickly learn link parameters so as to efficiently track the optimal path.

The design of such routing policies is often referred to as an online shortest path routing problem in the literature \cite{awerbuch2004adaptive,gyorgy2006adaptive,Gyoergy2007,He2013,brun2016big}, and is a particular instance of a combinatorial Multi-Armed Bandit (combinatorial MAB) problem as introduced in \cite{Cesa-Bianchi2012}. In this paper, we study the {\it stochastic} version of this problem. More precisely, we consider a network, in which the transmission of a packet on a given link is successful with an unknown but fixed probability.
A packet is sent on a given link repeatedly until the transmission is successful; the number of time slots to complete the transmission is referred to as the \emph{delay} on this link.
We wish to route $N$ packets from a given source to a given destination in a minimum amount of time. A routing policy selects a path to the destination on a packet-by-packet basis. The path selection can be done at the source (source routing), or in the network as the packet progresses towards the destination (hop-by-hop routing). In the case of source routing, some feedback is available when the packet reaches the destination. This feedback can be either the end-to-end delay, or the delays on each link on the path from source to destination. In the MAB literature, the former type of feedback is referred to as {\it bandit} feedback, whereas the latter is called {\it semi-bandit} feedback. The routing policy then selects the path for the next packet based on the feedback gathered from previously transmitted packets. In the case of hop-by-hop routing, routing decisions are taken for each transmission, and the packet is sent over a link selected based on all transmission successes and failures observed so far (for the current packet, and all previously sent packets) on the various links.

The performance of a routing policy is assessed through its expected total delay, i.e., the expected time required to send all $N$ packets to the destination. Equivalently, it can be measured through the notion of {\it regret}, defined as the difference between the expected total delay under the policy considered and the expected total delay of an oracle policy that would be aware of all link parameters, and would hence always send the packets on the best path. Regret conveniently quantifies the loss in performance due to the fact that link parameters are initially unknown and need to be learnt.

In this paper, we first address two fundamental questions: (i) what is the benefit of allowing routing decisions at every node, rather than only at the source? and (ii) what is the added value of feeding back the observed delay for every link that a packet has traversed compared to only observing the end-to-end delay?\footnote{The effect of different feedback in the adversarial setting was studied in, e.g., \cite{gyorgy2006adaptive,Gyoergy2007}.}  To answer these questions, we derive tight regret lower bounds satisfied by any routing policy in the different scenarios, depending on where routing decisions are made and what information is available to the decision-maker when making these decisions. By comparing the different lower bounds, we are able to quantify the value of having semi-bandit feedback rather than bandit feedback, and the improvements that can possibly be achieved by taking routing decisions hop by hop. We then propose routing policies in the semi-bandit feedback setting, and show that these policies outperform state-of-the-art online shortest path routing algorithms. More precisely, our contributions are the following:

\medskip
\noindent
\underline{1. Regret lower bounds.}  We derive tight asymptotic (when $N$ grows large) regret lower bounds. The two first bounds concern source routing policies under bandit and semi-bandit feedback, respectively, whereas the third bound is satisfied by any hop-by-hop routing policy. As we shall see later, these bounds are tight in the sense that there exist policies that achieve them. As it turns out, the regret lower bounds for source routing policies with semi-bandit feedback and that for hop-by-hop routing policies are identical, indicating that taking routing decisions hop by hop does not bring any advantage. On the contrary, the regret lower bounds for source routing policies with bandit and semi-bandit feedback can be significantly different, illustrating the importance of having information about per-link delays.

\medskip
\noindent
\underline{2. Routing policies.} In the case of semi-bandit feedback, we propose three online source routing policies, namely \mbox{\textsc{GeoCombUCB-1}}, \mbox{\textsc{GeoCombUCB-2}}, and KL-SR  (KL-based Source-Routing). \mbox{\textsc{Geo}} refers to the fact that the delay on a given link is geometrically distributed, \mbox{\textsc{Comb}} stands for combinatorial, and \mbox{\textsc{UCB}} (Upper Confidence Bound) indicates that these policies are based on the same ``optimism in face of uncertainty" principle as the celebrated UCB algorithm designed for classical MAB problems \cite{Auer2002}. KL-SR already appears in the conference version of this paper \cite{ZouACC2014}. Here we improve its regret analysis, and show that the latter scales at most as {\color{black}${\cal O}(|E|H\Delta_{\min}^{-1}\theta_{\min}^{-2}\log(N))$,} \footnote{This improves over the regret upper bound scaling as ${\cal O}(\Delta_{\max}|E|H^3\Delta_{\min}^{-1}\theta_{\min}^{-3}\log(N))$ derived in \cite{ZouACC2014}, where $\Delta_{\max}$ denotes the maximal gap between the average end-to-end delays of the optimal and of a sub-optimal path.} where $H$ denotes the length (number of links) of the longest path in the network from the source to the destination, $\theta_{\min}$ is the success transmission probability of the link with the worst quality, and $\Delta_{\min}$ is the minimal gap between the average end-to-end delays of the optimal and of a sub-optimal path (formal definitions of $\theta_{\min}$ and $\Delta_{\min}$ are provided in Section \ref{sec:model_NetworkModel}). We further show that the regret under \mbox{\textsc{GeoCombUCB-1}} and \mbox{\textsc{GeoCombUCB-2}} scales at most as ${\cal  O}(|E|\sqrt{H}\Delta_{\min}^{-1}\theta_{\min}^{-2}\log(N))$. The tradeoff between computational complexity and performance (regret) of online routing policies is certainly hard to characterize, but our policies provide a first insight into such a trade-off. Furthermore, they exhibit better regret upper bounds than that of the CUCB (Combinatorial UCB) algorithm \cite{chen2013combinatorial_icml}, which is, to our knowledge, the state-of-the-art {\color{black}algorithm for stochastic online shortest path routing}. Furthermore, we conduct numerical experiments, showing that our routing policies perform significantly better than CUCB.
{\color{black} The Thompson Sampling (TS) algorithm of~\cite{gopalan2014thompson} is applicable to the shortest path problem, but its analysis for general topologies is an  open problem. While TS performs slightly better than our algorithms on average, its regret sometimes has a large variance according to our experiments.} The regret guarantees of various algorithms, and their computational complexity are summarized in Table~\ref{table:comparison_regret}.

\begin{table}
\centering
\footnotesize
\begin{tabular}[b]{|c|c|c|}
\hline
\textbf{Algorithm} &  \textbf{Regret} & \textbf{Complexity} \\ \hline
\textsc{CUCB} \cite{chen2013combinatorial_icml} & ${\cal O}\left(\frac{|E|H}{\Delta_{\min}\theta_{\min}^3} \log(N)\right)$ & ${\cal O}(|V||E|)$ \\ \hline
\textsc{GeoCombUCB-1} %(Theorem \ref{thm:regret_geocombucb})
& ${\cal O}\left(\frac{|E|\sqrt{H}}{\Delta_{\min} \theta_{\min}^2}  \log(N)\right)$ & ${\cal O}(|\Pcal|)$ \\ \hline
\textsc{GeoCombUCB-2} %(Theorem \ref{thm:regret_geocombucb})
& ${\cal O}\left(\frac{|E|\sqrt{H}}{\Delta_{\min} \theta_{\min}^2}  \log(N)\right)$ & ${\cal O}(|\Pcal|)$ \\ \hline
\textsc{KL-SR} %(Theorem \ref{thm:regret_geocombucb_2})
& ${\cal O}\left(\frac{|E|H}{\Delta_{\min} \theta_{\min}^2} \log(N)\right)$ & ${\cal O}(|V||E|)$ \\ \hline
\end{tabular}
\caption{Comparison of various algorithms for shortest path routing under semi-bandit feedback.}
\label{table:comparison_regret}
\normalsize
\end{table}

\medskip
The remaining of the paper is organized as follows. In Section~\ref{sec:relatedWork} we review the literature related to MAB problems and to online shortest path problems. In Section~\ref{sec:model}, we introduce the network model and formulate our online routing problem. Fundamental performance limits (regret lower bounds) are derived in Section~\ref{sec:lowerBound}. We propose online routing algorithms and evaluate their performance in Section~\ref{sec:algorithm}. Finally, Section~\ref{sec:conclusion} concludes the paper and provides future research directions. All the proofs are presented in the Appendix.

\section{Related Work}
\label{sec:relatedWork}

Stochastic MAB problems have been introduced by Robbins \cite{Robbins1952}. In the classical setting, in each round, a decision maker pulls an arm from a set of available arms and observes a realization of the corresponding reward, whose distribution is unknown. The performance of a policy is measured through its regret, defined as the difference between its expected total reward and the optimal reward the decision maker could collect if she knew the reward distributions of all arms.
The goal is to find an optimal policy with the smallest regret. This classical stochastic MAB problem was solved by Lai and Robbins in their seminal paper \cite{Lai1985}, where they derived the asymptotic (when the time horizon is large) lower bound of regret satisfied by any algorithm, and proposed an optimal algorithm that matches the lower bound.

Online shortest path routing problems fall into the class of combinatorial MAB problems. In these MAB problems, arms are subsets of a set of basic actions (in routing problems, a basic action corresponds to a link), and most existing studies concern the adversarial setting where the successive rewards of each arm are arbitrary, see e.g. \cite{Cesa-Bianchi2012,Audibert2014,Bubeck2012towards,neu2013efficient} for algorithms for generic combinatorial problems, and \cite{awerbuch2004adaptive,Gyoergy2007} for efficient algorithms for routing problems. Stochastic combinatorial MAB problems have received little attention so far. Usually they are investigated in the semi-bandit feedback setting  \cite{Gai2012,chen2013combinatorial_icml,kveton2014tight,combes2015stochastic}. Some papers deal with problems where the set of arms exhibits very specific structures, such as fixed-size sets~\cite{Anantharam1987}, matroid~\cite{kveton2014matroid}, and permutations~\cite{gai2010learning}.

In the case of online shortest path routing problems, as a particular instance of a combinatorial MAB, one could think of modeling each path as an arm, and applying sequential arm selection policies as if arms would yield independent rewards. Such policies would have a regret scaling as $|{\cal P}|\log(N)$ where $|{\cal P}|$ denotes the number of possible paths from the source to the destination. However, since $|{\cal P}|$ grows exponentially with the length $H$ of the paths, treating paths as independent arms would lead to a prohibitive regret. In contrast to classical MAB in \cite{Lai1985} where the random rewards from various arms are \textit{independent}, in online routing problems, the end-to-end delays (i.e., the rewards) of the various paths are inherently correlated, since paths may share the same links. It may then be crucial to exploit these correlations, i.e., the structure of the problem, to design efficient routing algorithms which in turn may have a regret scaling as $C\log(N)$ where $C$ is much smaller than $|{\cal P}|$.

Next we summarize existing results for generic stochastic combinatorial bandits that could be applied to online shortest path routing. In \cite{chen2013combinatorial_icml}, the authors present CUCB, an algorithm for generic stochastic combinatorial MAB problems under semi-bandit feedback. When applied to the online routing problem, the best regret upper bound for CUCB presented in \cite{chen2013combinatorial_icml} scales as ${\cal O}(\frac{|E|H}{\Delta_{\min}\theta_{\min}^3}\log(N))$ (see Appendix \ref{sec:CUCB_regret} for details). % (see Appendix J in \cite{routing_paper_TechRep} for details).

This upper bound constitutes the best existing result for our problem, where the delay on each link is geometrically distributed. It is important to note that most proposed algorithms for combinatorial bandits \cite{Gai2012,kveton2014tight,combes2015stochastic} deal with bounded rewards, i.e., here bounded delays, and are not applicable to geometrically distributed delays. In \cite{kveton2014tight}, the authors consider the case where the rewards of basic actions (here links) can be arbitrarily correlated and bounded, and show that the regret under CUCB is ${\cal O}(\frac{|E|H}{\Delta_{\min}}\log(N))$. They also prove that this regret scaling has order-optimal regret in terms of $|E|$ and $H$\footnote{ A policy $\pi$ is order-optimal in terms of $|E|$ and $H$, if it satisfies the following: for all problem instances, $R^{\pi}(N)={\cal O}(C_1 g(|E|,H)\log(N))$ with $C_1$ independent of $|E|$, $H$, and $N$, and there exists a problem instance and a constant $C_2>0$, independent of $|E|$, $H$, and $N$, such that $\liminf_{N\to\infty} R^{\pi'}(N)/\log(N) \ge C_2g(|E|,H)$ for all uniformly good algorithm $\pi'$.}. In other words, the dependence of their regret upper bound on $|E|$ and $H$ cannot be improved {\color{black} in general}. This order-optimality does not contradict our regret upper bound (scaling as ${\cal O}(\frac{|E|\sqrt{H}}{\Delta_{\min}}\log(N))$),  because \cite{kveton2014tight} considers possibly dependent delays across links. Interestingly, to prove that a regret of ${\cal O}(\frac{|E|H}{\Delta_{\min}}\log(N))$ cannot be beaten, they artificially create an instance of the problem where the rewards of the basic actions of the same arm are identical. In other words, they consider a classical bandit problem where the rewards of the various arms are either 0 or equal to $H$. This clearly highlights the fact that the approach of \cite{kveton2014tight} cannot be directly applied to our routing problem where delays are unbounded. For bounded rewards, the results of \cite{kveton2014tight} have been recently improved in \cite{combes2015stochastic} when the rewards are \emph{independent} across basic actions (links). There, the authors propose an algorithm whose regret scales at most as ${\cal O}(\frac{|E|\sqrt{H}}{\Delta_{\min}}\log(N))$. Wen et al.~\cite{wen2015efficient} study combinatorial problems under semi-bandit feedback and provide algorithms with ${\cal O}(\sqrt{N})$ regret. Gopalan et al.~\cite{gopalan2014thompson} study TS \cite{thompson1933likelihood} for learning problems with complex arms and provide implicit regret upper bounds with ${\cal O}(\log(N))$ regret.

Stochastic online shortest path routing problems have been addressed in  \cite{Liu2012,He2013,tehrani2013distributed}. Liu and Zhao \cite{Liu2012} consider routing with bandit (end-to-end) feedback and propose a forced-exploration algorithm with ${\cal O}(|E|^3H\log(N))$ regret in which a random barycentric spanner\footnote{A barycentric spanner is a set of paths from which the delay
of all other paths can be computed as its linear combination with coefficients in $[-1,1]$ \cite{awerbuch2004adaptive}.} path is chosen for exploration.
He et al.~\cite{He2013} consider routing under semi-bandit feedback, where the source chooses a path for routing and a possibly different path for probing. Our model coincides with the coupled probing/routing case in their paper, for which they derive an asymptotic lower bound on the regret growing logarithmically with time. As we shall see later, their lower bound is not tight.

Finally, it is worth noting that the papers cited above considered source-routing only. To the best of our knowledge, this paper is the first to consider online routing problems  with hop-by-hop decisions. Such a problem can be formulated as a classical Markov Decision Process (MDP), in which the states are the packet locations and the actions are the outgoing links of each node. However, most studies consider MDP problems under stricter assumptions than ours and/or targeted different performance measures. Burnetas and Katehakis~\cite{Burnetas1997} derive the asymptotic lower bound on the regret and propose an optimal index policy. Their result can be applied only to the so-called ergodic MDP~\cite{Puterman2005}, where the induced Markov chain by any policy is irreducible and consists of a single recurrent class. In hop-by-hop routing, however, the policy that routes packets on a fixed path results in a Markov chain with reducible states that are not in the chosen path. %Algorithms for general MDPs with logarithmic regret were also proposed in, e.g., \cite{jaksch2010,Filippi2010}.
{\color{black} \cite{jaksch2010,Filippi2010} study general MDPs and present algorithms with finite-time regret upper bounds scaling as ${\cal O}(\log(T))$. Nevertheless, these algorithms perform badly when applied to hop-by-hop routing due to loose confidence intervals. \cite{jaksch2010} also presents non-asymptotic, but problem independent (minimax) regret lower bounds scaling as $\Omega(\sqrt{T})$. This latter bound does not contradict our problem-dependent lower bounds that grow logarithmically.

% both worst-case bounds for fixed $T$ and problem-dependent logarithmic bounds, both of these holding for finite $T$. , and only asymptotic regret upper bounds are known.
}

\begin{comment}
Lastly, MDP under bandit feedback has also been studied under the PAC (Probably Approximately Correct) model~\cite{Even-Dar2002,Strehl2006,Strehl2009}, where basic algorithms were derived that find, with high probability, a near-optimal action at each state for any sample path of the MDP after a specified learning period.
The corresponding regret bound can be shown to scale logarithmically with time. However, the regret is measured by the expected discounted total reward with a discount factor strictly smaller than one.
\end{comment}
%In summary, we obtain (for the first time) a tight asymptotic regret lower bound for source routing with bandit and semi-bandit feedback, and for hop-by-hop routing.
%Moreover, we derive several practical routing algorithms that achieve better regrets than existing algorithms both in theory and in numerical experiments as will be shown in Section~\ref{sec:algorithm}.

\section{Online shortest path Routing Problems}
\label{sec:model}
\subsection{Network Model}
\label{sec:model_NetworkModel}
The network is modeled as a directed graph $G=(V,E)$ where $V$ is the set of nodes and $E$ is the set of links.
Each link $i\in E$ may, for example, represent an unreliable wireless link. Without loss of generality, we assume that time is slotted and that one slot corresponds to the time to send a packet over a single link. At time $t$, $X_i(t)$ is a binary random variable indicating whether a transmission on link $i$ at time $t$ is  successful. $(X_i(t))_{t\ge 1}$ is a sequence of i.i.d. Bernoulli variables with initially unknown mean $\theta_i$.
Hence if a packet is sent on link $i$ repeatedly until the transmission is successful,  the time to complete the transmission (referred to as the delay on link $i$) is geometrically distributed with mean $1 / \theta_i$. Let $\theta=(\theta_i,i\in E)$ be the vector representing the packet successful transmission probabilities on the various links.
We consider a single source-destination pair $(s,d)\in V^2$, and denote by ${\cal P}\subseteq \{0,1\}^{|E|}$ the set of loop-free paths from $s$ to $d$ in $G$, where each path $p\in \Pcal$ is a $|E|$-dimensional binary vector; for any $i\in E$, $p_i=1$ if and only if $i$ belongs to $p$. Let $H$ denote the maximum length of the paths in ${\cal P}$, i.e., $H=\max_{p\in\Pcal}\sum_{i\in E}p_i$.
For brevity, in what follows, for any binary vector $z$, we write $i\in z$ to denote $z_i=1$. Moreover, we use the convention that $z^{-1}=(z_i^{-1})_i$.

For any path $p$, $D_\theta(p)=\sum_{i\in p}{ \frac{1}{\theta_i} }$ is the average packet delay through path $p$ given link success rates $\theta$. The path with minimal delay is:
$p^\star \in \arg\min_{p\in {\cal P}} D_\theta(p).$
Moreover, for any path $p\in \Pcal$, we define $\Delta_p=D_{\theta}(p)-D_{\theta}(p^\star)=(p-p^\star)^\top\theta^{-1}$. Let $\Delta_{\min}=\min_{\Delta_p\ne 0} \Delta_p$.
We let $\theta_{\min}=\min_{i\in E}\theta_i$ and assume that $\theta_{\min}>0$. Finally define $D^\star=D_\theta({p^\star})$ and {\color{black}$D^+= \max_{p \in {\cal P}} D_\theta({p})$} the delays of the shortest and longest paths, respectively.

The analysis presented in this paper can be easily extended to more general link models, provided that the (single-link) delay distributions are taken within one-parameter exponential families of distributions.
%Please refer to the remarks after Theorem~\ref{thm:SourceAggregate} and Theorem~\ref{thm:SourceSemi-bandit} for details.

\subsection{Online Routing Policies and Feedback}

We assume that the source is fully backlogged (i.e., it always has packets to send), and that the parameter $\theta$ is initially unknown. Packets are sent successively  from $s$ to $d$ over various paths, and the outcome of each packet transmission is used to estimate $\theta$, and in turn to learn the path $p^\star$ with the minimum average delay. After a packet is sent, we assume that the source gathers feedback from the network (essentially per-link or end-to-end delays) before sending the next packet.

Our objective is to design and analyze online routing policies, i.e., policies that take routing decisions based on the feedback received for the packets previously sent.

We consider and compare three different types of online routing policies, depending (i) on where routing decisions are taken (at the source or at each node), and (ii) on the received feedback (per-link or end-to-end path delay). Table \ref{table:policy_sets} lists different policy sets for the three types of online routing policies considered.

\begin{table}
\centering
\footnotesize
\begin{tabular}[b]{|c|c|c|}
\hline
\textbf{Policy Set} &  \textbf{Routing Type} & \textbf{Feedback} \\ \hline
$\Pi_1$ & Source-routing & Bandit \\ \hline
$\Pi_2$ & Source-routing & Semi-bandit \\ \hline
$\Pi_3$ & Hop-by-hop & Semi-bandit \\ \hline
\end{tabular}
\caption{Various policy sets for online shortest path routing.}
\label{table:policy_sets}
\normalsize
\end{table}

\begin{itemize}
\item \underline{Policy Set $\Pi_1$:} The path used by a packet is determined at the source based on the observed end-to-end delays for previous packets. More precisely, for the $n$-th packet, let $p^\pi(n)$ be the path selected under policy $\pi$, and let $D^\pi(n)$ denote the corresponding end-to-end delay. Then $p^\pi(n)$ depends on $p^\pi(1),\ldots,p^\pi(n-1), D^\pi(1),\ldots,D^\pi(n-1)$.
\item \underline{Policy Set $\Pi_2$:} The path used by a packet is determined at the source based on the observed per-link delays for previous packets. In other words, under policy $\pi$, $p^\pi(n)$ depends on $p^\pi(1),\ldots,p^\pi(n-1), (d_i^\pi(1), i\in p^\pi(1)),\ldots,(d_i^\pi(n-1), i\in p^\pi(n-1))$, where $d^\pi_i(k)$ is the  delay experienced on link $i$ for the $k$-th packet (if this packet uses link $i$ at all).
\item \underline{Policy Set $\Pi_3$:} Routing decisions are taken at each node in an adaptive manner. At a given time $t$, the packet is sent over a link selected based on all successes and failures observed on the various links before time $t$.
\end{itemize}

%\begin{itemize}
%\item {\it Source Routing with Bandit Feedback.} The path used by a packet is determined at the source based on the observed end-to-end delays for previous packets. More precisely, for the $n$-th packet, let $p^\pi(n)$ be the path selected under policy $\pi$, and let $D^\pi(n)$ denote the corresponding end-to-end delay. Then $p^\pi(n)$ depends on $p^\pi(1),\ldots,p^\pi(n-1), D^\pi(1),\ldots,D^\pi(n-1)$. We denote by $\Pi_1$ the set of such policies.
%\item {\it Source Routing with Semi-bandit Feedback.} The path used by a packet is determined at the source based on the observed per-link delays for previous packets. In other words, under policy $\pi$, $p^\pi(n)$ depends on $p^\pi(1),\ldots,p^\pi(n-1), (d_i^\pi(1), i\in p^\pi(1)),\ldots,(d_i^\pi(n-1), i\in p^\pi(n-1))$, where $d^\pi_i(k)$ is the  delay experienced on link $i$ for the $k$-th packet (if this packet uses link $i$ at all). We denote by $\Pi_2$ the set of such policies.
%\item {\it Hop-by-hop Routing.} Routing decisions are taken at each node in an adaptive manner. At a given time $t$, the packet is sent over a link selected based on all successes and failures observed on the various links before time $t$. Let $\Pi_3$ denote the set of hop-by-hop routing policies.
%\end{itemize}

In the case of source-routing policies (in $\Pi_1\cup\Pi_2$), if a transmission on a given link fails, the packet is retransmitted on the same link until it is successfully received (per-link delays are geometric random variables). On the contrary, in the case of hop-by-hop routing policies (in $\Pi_3$), the routing decisions at a given node can be adapted to the observed failures on a given link. For example, if transmission attempts on a given link failed, one may well decide to switch link and select a different next-hop node.

\subsection{Performance Metrics and Objectives}

\subsubsection{Regret}

Under any reasonably smart routing policy, the parameter $\theta$ will eventually be estimated accurately and the minimum delay path will be discovered with high probability after sending a large number of packets.
Hence, to quantify the performance of a routing policy, we examine its transient behavior. More precisely, we use the notion of {\it regret}, a performance metric often used
%in online stochastic optimization, and
in MAB literature~\cite{Lai1985}. The regret $R^\pi(N)$ of policy $\pi$ up to the $N$-th packet is the expected difference of delays for the first $N$ packets under $\pi$ and under the policy that always selects the best path $p^\star$ for transmission:
$$
R^\pi(N) := \EE \left[ \sum_{n=1}^{N} D^\pi(n) \right]
  - N D_\theta (p^\star),
$$
where $D^\pi(n)$ denotes the end-to-end delay of the $n$-th packet under policy $\pi$ and the expectation $\EE[\cdot]$ is taken with respect to the random transmission outcomes and possible randomization in the policy $\pi$. The regret quantifies the performance loss due to the need to explore sub-optimal paths to learn the path with minimum delay.

\subsubsection{Objectives}

The goal is to design online routing policies in $\Pi_1$, $\Pi_2$, and $\Pi_3$ that minimize regret over the first $N$ packets.
% This can be formulated as an online stochastic optimization problem.
% (see \cite{bubek} for an introduction).
% Bubek's lecture notes actually only cover the adversary type and only a section is devoted to the basics of the stochastic version (Lai and Robbins, and Auer's UCB policy.)
%
% this problem is often referred to as a combinatorial bandit problem \cite{gyorgy2007,CesaBianchi2012}.
As it turns out, there are policies in any $\Pi_j$, $j=1,2,3$, whose regrets scale as ${\cal O}(\log(N))$ when $N$ grows large, and no policy can have a regret scaling as $o(\log(N))$.
%~\footnote{This is not obvious for hop-by-hop routing.
% But in source routing, we can model each path as an arm and the above results follow by arguments from classical MAB bounds~\cite{Lai1985,Auer2002} based on that the joint distribution over all arms are i.i.d. over time.}
% A reviewer points out that this result is obvious for source routing case, and we should not use the word "as it turns out"...
%

Our objective is to derive, for each $j=1,2,3$, an asymptotic regret lower bound $c_j(\theta)\log(N)$ for policies in $\Pi_j$, and then propose simple policies whose regret upper bounds asymptotically approach that of the \emph{optimal} algorithm, i.e., an algorithm whose regret matches the lower bound in $\Pi_j$. As we shall discuss later, there exists an algorithm whose regret asymptotically matches these lower bound. Therefore, by comparing $c_1(\theta)$, $c_2(\theta)$, and $c_3(\theta)$, we can quantify the potential performance improvements taking routing decisions at each hop rather than at the source only, and observing per-link delays rather than end-to-end delays.

\section{Fundamental Performance Limits}
\label{sec:lowerBound}

In this section, we provide fundamental performance limits satisfied by {\it any} online routing policy in $\Pi_1$, $\Pi_2$, or $\Pi_3$. Specifically, we derive asymptotic (when $N$ grows large) regret lower bounds for our three types of policies. These bounds are obtained exploiting some results and techniques used in the control of Markov chains~\cite{Graves1997}, and they are {\it tight} in the sense that there exist algorithms achieving these performance limits.

\subsection{Regret Lower Bounds}

We restrict our attention to the so-called {\it uniformly good} policies, under which the number of times sub-optimal paths are selected until the transmission of the $n$-th packet is $o(n^\alpha)$ when $n\to\infty$ for any $\alpha >0$ and for all $\theta$. We know from~\cite[Theorem~2]{Graves1997} that such policies exist.

\subsubsection{Source-Routing with Bandit Feedback}

Denote by $\psi_\theta^p(k)$ the probability that the delay of a packet sent on path $p$ is $k$ slots, and by $h(p)$ the length (or number of links) of path $p$.
The end-to-end delay is the sum of several independent random geometric variables. If we assume that $\theta_i\neq \theta_j$ for $i\neq j$, we have~\cite{Sen1999}, for all $k\ge h(p)$,
$$
  \psi_\theta^{p}(k) = \sum_{i \in p}
  \biggl( \prod_{j\in p, j\neq i} \dfrac{\theta_j}{\theta_j - \theta_i} \biggr)
  \theta_i  (1-\theta_i)^{k-1},
$$
i.e., the path delay distribution is a weighted average of the individual link delay distributions where the weights can be negative but always sum to one.

The next theorem provides the fundamental performance limit of online routing policies in $\Pi_1$.

\begin{theorem}
\label{thm:SourceAggregate}
For all $\theta$ and for any uniformly good policy $\pi\in \Pi_1$, % we have
%\begin{align*}
   $ \liminf_{N \rightarrow \infty}
    \frac{R^{\pi}(N)}{\log(N)} \geq c_1(\theta),$
%\end{align*}
where $c_1(\theta)$ is the infimum of the following optimization problem:
\begin{align}
\inf_{x\ge 0}& \;\; \sum_{p \in \Pcal} x_{p} \Delta_p
\label{eq:SourceRoutingAgg}
\\ %(D_\theta(p)-D_\theta(p^\star))\\
\hbox{subject to:} & \;\; %c_{p} \geq 0, \forall p\in {\cal P}, \\
 \inf_{\lambda \in B_1(\theta)} \sum_{p \neq p^\star} x_{p} \sum_{k = h(p)}^{\infty}
  \psi_\theta^{p}(k) \log \dfrac{\psi_\theta^{p}(k)}{\psi_\lambda^{p}(k)}
  \geq 1, \nonumber
  \end{align}
with
\small
\begin{align*}
B_1(\theta) = \Bigl\{ \lambda:\{\lambda_i, i\in p^\star\}= \{ \theta_i, i \in p^\star\}, \;
\min_{p\in {\cal P}} D_\lambda(p)  < D_\lambda(p^\star) \Big\}.
\end{align*}
\normalsize
\end{theorem}

\medskip
The variables $x_p, p\in \Pcal$ solving (\ref{eq:SourceRoutingAgg}) have the following interpretation: for $p\neq p^\star$, $x_p\log(N)$ is the asymptotic number of packets that needs to be sent (up to the $N$-th packet) on sub-optimal path $p$ under optimal routing strategies in $\Pi_1$. Hence, $x_p$ determines the optimal rate of {\it exploration} of sub-optimal path $p$. $B_1(\theta)$ is the set of {\it bad} network parameters: if $\lambda \in B_1(\theta)$, then the end-to-end delay distribution along the optimal path $p^\star$ is the same under $\theta$ or $\lambda$ (hence by observing the end-to-end delay on path $p^\star$, we cannot distinguish $\lambda$ or $\theta$), and $p^\star$ is not optimal under $\lambda$.

It is important to observe that in the definition of $B_1(\theta)$, the equality $\{\lambda_i, i\in p^\star\} = \{ \theta_i, i \in p^\star\}$ is a set equality, i.e., order does not matter (e.g., if $p^\star=\{1,2\}$, the equality means that either $\lambda_1=\theta_1,\lambda_2=\theta_2$ or $\lambda_1=\theta_2,\lambda_2=\theta_1$).

\subsubsection{Source-Routing with Semi-Bandit (Per-Link) Feedback}

We now consider routing policies in $\Pi_2$ that make decisions at the source, but have information on the individual link delays. Let $\klg(u,v)$ denote the
%Kullback-Leibler
KL divergence number between two geometric random variables with parameters $u$ and $v$:
\begin{align*}
\klg(u,v) := \sum_{k \geq 1} u (1-u)^{k-1}\log \dfrac{u (1-u)^{k-1}}{v (1-v)^{k-1}}.
\end{align*}

\begin{theorem}
\label{thm:SourceSemi-bandit}
For all $\theta$ and for any uniformly good policy $\pi\in \Pi_2$, % we have
%  \begin{align*}
    $\liminf_{N \rightarrow \infty}
    \frac{R^{\pi}(N)}{\log(N)} \geq c_2(\theta)$,
 % \end{align*}
where $c_2(\theta)$ is the infimum of the following optimization problem:
\begin{align}
\inf_{x\ge 0} & \;\; \sum_{p \in \Pcal} x_{p} \Delta_p
\label{eq:SourceRoutingDetail}  \\%\;\;\;\;\;
\hbox{subject to:} & \;\;
\inf_{\lambda \in B_2(\theta)} \sum_{p\neq p^\star} x_{p} \sum_{i \in p}\klg(\theta_i, \lambda_i) \geq 1, \nonumber
\end{align}
with
$$
B_2(\theta) =\{ \lambda : \lambda_i = \theta_i, \; \forall i \in p^\star,
\min_{p\in {\cal P}} D_\lambda(p) < D_\lambda(p^\star)\}.
$$
\end{theorem}
%\begin{remark}
%Theorem~\ref{thm:SourceSemi-bandit} holds for any parametric link delay distribution by replacing $\klg(\cdot,\cdot)$ with the corresponding KL divergence number.
%\end{remark}
% The path that has hop-count larger than or equal to the average delay of the shortest path does not appear in the optimization problem.
% The infimum over the empty set is infinity..
\medskip
The variables $x_p,p\in\Pcal$ solving (\ref{eq:SourceRoutingDetail}) have the same interpretation as that given previously in the case of bandit feedback. Again $B_2(\theta)$ is the set of parameters $\lambda$ such that the distributions of link delays along the optimal path are the same under $\theta$ and $\lambda$, and $p^\star$ is not the optimal path under $\lambda$. The slight difference between the definitions of $B_1(\theta)$ and $B_2(\theta)$ comes from the difference of feedback (bandit vs. semi-bandit). It is also noted that $B_2(\theta)\subset B_1(\theta)$.  We stress that by \cite[Theorem~2]{Graves1997}, the asymptotic regret lower bounds of Theorems \ref{thm:SourceAggregate}-\ref{thm:SourceSemi-bandit} are tight, namely there exists policies that achieve these regret bounds.

\begin{remark} Of course, we know that $c_1(\theta)\ge c_2(\theta)$, since the lower bounds we derive are tight and getting per-link delay feedback can be exploited to design smarter routing policies than those we can devise using end-to-end delay feedback (i.e., $\Pi_1\subset \Pi_2$).
%Note that the data processing inequality~\cite[Theorem 1]{vanErven2014} implies that for any $p\neq p^\star$:
%\begin{align*}
 % \sum_{k = h(p)}^{\infty}
 % \psi_\theta^{p}(k)\log \dfrac{\psi_\theta^{p}(k)}{\psi_\lambda^{p}(k)}
 % \leq \sum_{i \in p} \klg(\theta_i, \lambda_i),
 % \;\; \forall \theta,\lambda,
%\end{align*}
%which together with $B_2(\theta)\subset B_1(\theta)$ implies that $c_1(\theta) \geq c_2(\theta)$. Namely, having semi-bandit feedback improves performance.
\end{remark}

\begin{remark}
The asymptotic lower bound proposed in~\cite{He2013} has a similar expression to ours, but the set $B_2(\theta)$ is replaced by
$
 B_2'(\theta) =  \bigcup_{i \in E} \{ \lambda: \lambda_j=\theta_j, \forall j \neq i,
  \min_{p\in {\cal P}} D_\lambda(p) < D_\lambda(p^\star) \}.
$
Note that $B_2'(\theta)\subset B_2(\theta)$, which implies that the lower bound derived in~\cite{He2013} is
smaller than ours. In other words, we propose a regret lower bound that improves that in\cite{He2013}. Furthermore,  our bound is tight (it cannot be improved further).
\end{remark}

The proof of Theorems \ref{thm:SourceAggregate} and \ref{thm:SourceSemi-bandit} leverage techniques from \cite{Graves1997} developed for the control of Markov chains, and are presented in Appendix \ref{sec:proofLowerBounds}. Theorem \ref{thm:SourceSemi-bandit} can be seen as a direct consequence of \cite[Theorem~1]{Graves1997} (the problem can be easily mapped to a controlled Markov chain). In contrast, the proof of Theorem \ref{thm:SourceAggregate} requires a more clever mapping due to the different nature of feedback. To prove Theorem \ref{thm:SourceAggregate}, we establish Lemma~\ref{lem:Geo_permut}, a property for geometric random variables.

\subsubsection{Hop-by-hop Routing}

Finally, we consider routing policies in $\Pi_3$. These policies are more involved to analyze as the routing choices may change at any intermediate node in the network, and they are also more complex to implement. Surprisingly, the next theorem states that the regret lower bound for hop-by-hop routing policies is the same as that derived for strategies in $\Pi_2$ (source-routing with semi-bandit feedback). In other words, we cannot improve the performance by taking routing decisions at each hop.

\medskip
\begin{theorem}
\label{thm:HbH}
For all $\theta$ and for any uniformly good rule $\pi\in \Pi_3$,
$
\liminf_{N \rightarrow \infty} \frac{R^{\pi}(N)}{\log(N)} \geq c_3(\theta)=c_2(\theta).
$
\end{theorem}
\medskip

The proof of Theorem \ref{thm:HbH} is more involved than those of previous theorems, since in the hop-by-hop case, the chosen path could change at intermediate nodes. To overcome this difficulty, we introduce another notion of regret corresponding to the achieved throughput (i.e., the number of packets successfully received by the destination per unit time), which we refer to as the {\it throughput regret}. %More precisely, $S^\pi(T)$ of $\pi$ over time horizon $T$ is:
%$
%S^\pi(T) := T\mu_\theta(p^\star)
%- \EE \left[ N^\pi(T) \right],
%$
%where $N^\pi(T)$ is the number of packets received up to time $T$ under policy $\pi$.
The proof uses the results of \cite{Graves1997} for throughput regret, but also relies on Lemma~\ref{lem:delayEqThroughput}, which provides an asymptotic relationship between $R^\pi(N)$ and the throughput regret.

As shown in~\cite[Theorem~2]{Graves1997}, the asymptotic regret lower bounds derived in Theorems \ref{thm:SourceAggregate}-\ref{thm:SourceSemi-bandit}-\ref{thm:HbH} are \emph{tight} in the sense that one can design actual routing policies achieving these regret bounds (although these policies might well be extremely complex to compute and impractical to implement).
Hence from the fact that $c_1(\theta)\ge c_2(\theta)=c_3(\theta)$, we conclude that:
\begin{itemize}
\item The optimal source-routing policy with semi-bandit feedback asymptotically achieves a lower regret than the optimal source-routing policy with bandit feedback;
\item The optimal hop-by-hop routing policy asymptotically obtains the same regret as the optimal source-routing policy with semi-bandit feedback.
\end{itemize}

\subsection{Numerical Example}

\begin{figure}[!th]
\begin{center}
\subfigure[]{
\includegraphics[scale=.4]{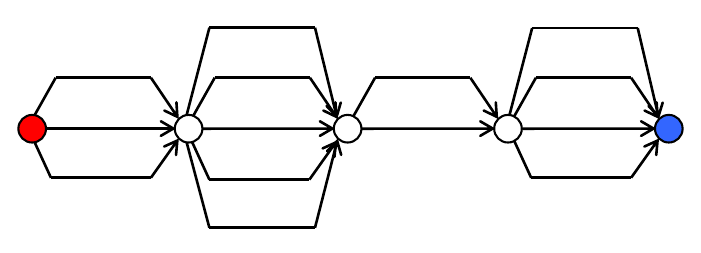}}
\subfigure[]{
\includegraphics[scale=.15]{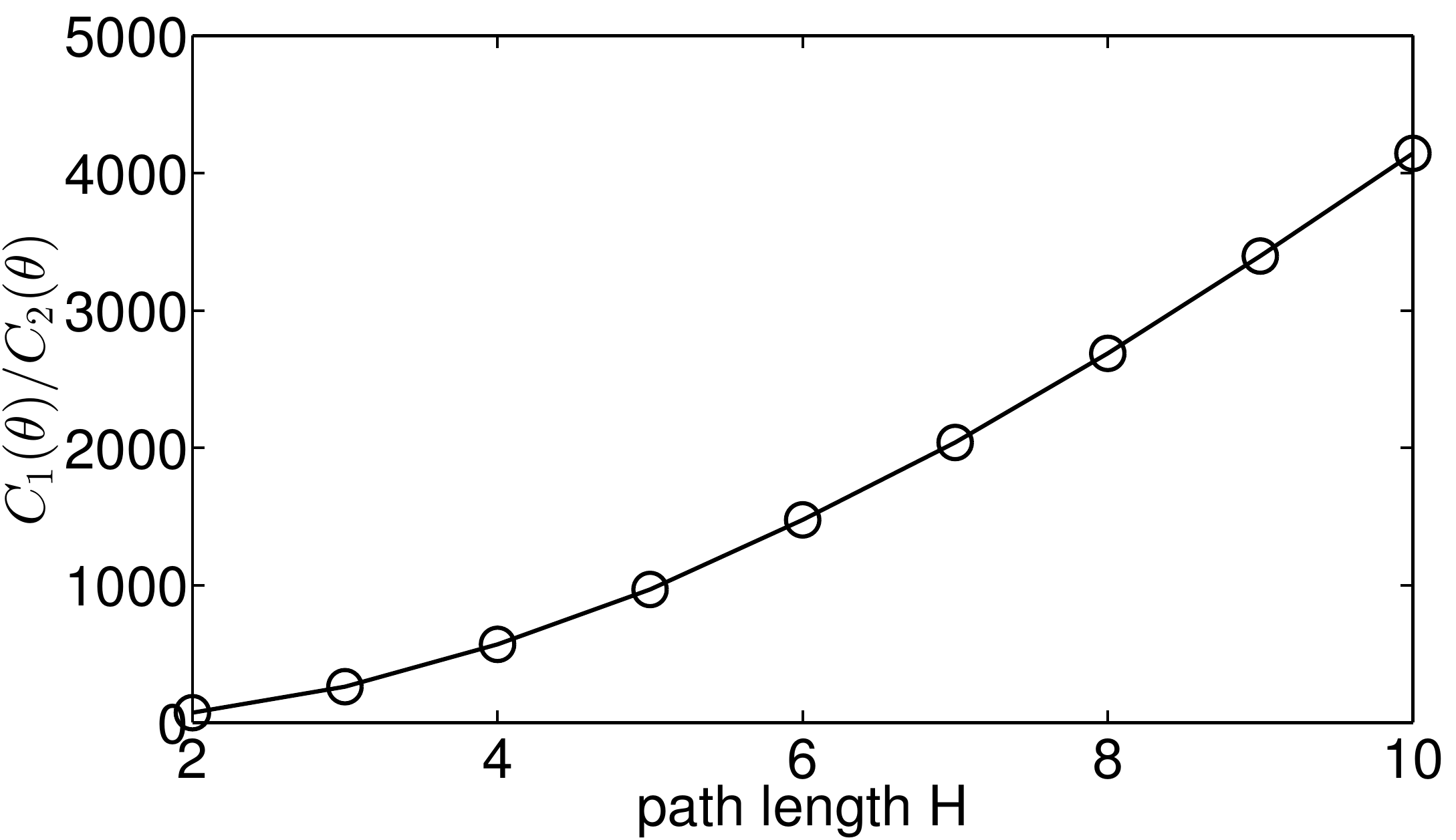}}
\end{center}
\caption{The line network: (a) Topology, (b) Semi-bandit vs. bandit feedback: lower bound on the average ratio between the two corresponding asymptotic regret lower bounds ($c_1(\theta)/c_2(\theta)$).}
\label{fig:LineNet}
\end{figure}

There are examples of network topologies where the above asymptotic regret lower bounds can be explicitly computed. One such example is the line network; see e.g. Figure~\ref{fig:LineNet}(a). Notice that in line networks, the optimal routing policy consists in selecting the best link in each hop. The following lemma is immediate:

\medskip
\begin{lemma}
\label{prop:LineNetwork_Ctheta}
For any line network with $H$ hops, we have:
\begin{align*}
c_1(\theta) & \ge  \sum_{i\notin p^\star}\dfrac{ \frac{1}{\theta_i} - \frac{1}{\theta_{\zeta(i)}} }
{ \max_{p:i\in p} \sum_{k = H}^{\infty} \psi_\theta^{p}(k)
  \log \frac{\psi_\theta^{p}(k)}{\psi_{\vartheta^i}^{p}(k)} },\\
c_2(\theta) &= c_3(\theta) = \sum_{i\notin p^\star}\dfrac{ {\frac{1}{\theta_i}}  - \frac{1}{\theta_{\zeta(i)}} }
{  \klg(\theta_i, \theta_{\zeta(i)}) },
\end{align*}
where $\zeta(i)$ is the best link on the same hop as link $i$ and $\vartheta^i$ is a vector of link parameters defined as $\vartheta_j^i=\theta_j$ if $j\neq i$, and $\vartheta_i^i=\theta_{\zeta(i)}$.
\end{lemma}

\begin{proposition}
\label{prop:LineNetwork_LB_example}
There exist problem instances in line networks, for which the regret of any uniformly good policy in $\Pi_2\cup\Pi_3$ is $\Omega\Bigl(\frac{|E|-H}{\Delta_{\min}\theta_{\min}^{2}}\log(N)\Big)$.
\end{proposition}

%{\color{blue}
%\begin{remark}
%As shown in Appendix C of \cite{routing_paper_TechRep}, for line networks, the above lemma implies that the regret of any uniformly good policy in $\Pi_2\cup\Pi_3$ is $\Omega\Bigl(\frac{|E|-H}{\Delta_{\min}\theta_{\min}^{2}}\log(N)\Big)$.
%\end{remark}
%}

\medskip
For line networks, both $c_1(\theta)$ and $c_2(\theta)$ scale linearly with the number of links in the network. In Figure \ref{fig:LineNet}(b), we plot the lower bound of the ratio $\frac{c_1(\theta)}{c_2(\theta)}$ (based on the previous lemma) averaged over various values of $\theta$ (we randomly generated $10^6$ link parameters $\theta$) as a function of the network size $H$ in a simple line network, which has two links in the first hop and one link in the rest of hops and hence $|E|=H+1$. These results suggest that collecting semi-bandit feedback (per-link delays) can significantly improve the performance of routing policies. The gain is significant even for fairly small networks  -- the regret is reduced by at least a factor 1500 on average in $6$-hop networks when collecting per-link delays.

\section{Routing Policies for Semi-bandit Feedback}
\label{sec:algorithm}

Theorems~\ref{thm:SourceAggregate}-\ref{thm:SourceSemi-bandit}-\ref{thm:HbH} indicate that within the first $N$ packets, the total amount of packets routed on a sub-optimal path $p$ should be of the order of $x^\star_p\log(N)$ where $x^\star_p$ is the optimal solution of the optimization problems in~\eqref{eq:SourceRoutingAgg}~and~\eqref{eq:SourceRoutingDetail}. In \cite{Graves1997}, the authors present policies that achieve the regret bounds of Theorems~\ref{thm:SourceAggregate}-\ref{thm:SourceSemi-bandit}-\ref{thm:HbH} (see \cite[Theorem~2]{Graves1997}). These policies suffer from two problems: firstly, they are computationally infeasible for large problems since their implementation involves solving in each round a semi-infinite linear program \cite{shapiro2009semi} similar to those providing the regret lower bounds (defined in \eqref{eq:SourceRoutingAgg}~and~\eqref{eq:SourceRoutingDetail}). Secondly, these policies have no finite-time performance guarantees, and numerical experiments suggest that their finite-time performance on typical problems is rather poor.

In this section, we present online routing policies for semi-bandit feedback, which are simple to implement, yet approach the performance limits identified in the previous section. We further analyze their regret, and show that they outperform existing algorithms. To present our policies, we introduce additional notations. Under a given policy, %we define $s_i(n)$ as the number of packets routed through link $i$ before the $n$-th packet is sent.
we let $t_i(n)$ be the total number of transmission attempts (including retransmissions) on link $i$ before the $n$-th packet is sent. We define $\hat{\theta}_i(n)$ the empirical success rate of link $i$ estimated over the transmissions of the {\color{black}first $(n-1)$ packets}. %; namely $\hat{\theta}_i(n)=s_i(n)/ t_i(n)$ if $t_i(n) > 0$ and $\hat\theta_i(n) = 0$ otherwise.
We define the corresponding vectors $t(n) = (t_i(n))_{i\in E}$ %, $s(n)=(s_i(n))_{i\in E}$,
and $\hat\theta(n) =(\hat\theta_i(n))_{i\in E}$.

Note that the proposed policies and regret analysis presented in this section directly apply for generic combinatorial optimization problems with linear objective function and geometrically distributed rewards.

%First we present \textsc{GeoCombUCB} and KL-SR policies that
%%The first policy, referred to as KL-SR (Kullback-Leibler-Source Routing),
%belong to $\Pi_2$ (routing decisions are taken at the source based on semi-bandit feedback). Then, we present KL-HHR (Kullback-Leibler-Hop-by-Hop Routing), which belongs to $\Pi_3$ (routing decisions are taken at each hop). These algorithms are simple index policies: to each path is attached an index, and packets are sent on the path with the current minimal index. The index of a given path is further defined through the indexes of its constituting links. The latter indexes are the same as those used in the KL-UCB algorithm \cite{Garivier2011}, an algorithm known to be asymptotically optimal for classical multi-armed bandit problems. We investigate the regret of these algorithms analytically and numerically; we show that they exhibit similar performance, and that they outperform existing algorithms. We also establish the asymptotic optimality of KL-SR in specific network topologies.

\subsection{Path and Link Indexes}

The proposed policies rely on indexes attached either to individual links or paths. Next we introduce three indexes used in our policies. They depend on the round, i.e., on the number $n$ of packets already sent, and on the estimated link parameters $\hat\theta(n)$. The three indexes and their properties (i.e., in which policy they are used, and how one can compute them) are summarized in Table~\ref{table:indexes}. Let $n\ge 1$ and assume that {\color{black}the $n$-th} packet is to be sent. The indexes are defined as follows.
% namely an index $\xi$ is a function of the form $\xi(n,\hat\theta(n))$. For brevity we omit the dependence on $\hat\theta(n)$ whenever there is no confusion.
%The first two indexes are \emph{path indexes}, that is, they are defined for each path $p\in \Pcal$.
%On the contrary, the last index is \emph{edge index} and is defined for each edge $i\in E$.
%Among these, one path index is explicitly defined with a closed-formed expression whereas the other two are defined as the optimal value of optimization problems. These latter indexes capture KL geometry of the problem and might be viewed as extensions of \textsc{KL-UCB} index in \cite{Garivier2011} for a combinatorial problem with geometric costs.

\begin{table}
\centering
\footnotesize
\begin{tabular}[b]{|c|c|c|c|}
\hline
\textbf{Index} &  \textbf{Type} & \textbf{Computation} & \textbf{Algorithm}\\ \hline
   $b_p$ & Path & Line search & \textsc{GeoCombUCB-1} \\ \hline
   $c_p$ & Path & Explicit & \textsc{GeoCombUCB-2}\\ \hline
   $\omega_i$ & Edge & Line search & \textsc{KL-SR} \\ \hline
%   $\gamma$ & Edge & Explicit &\textsc{GeoCombUCB-4} \\ \hline
\end{tabular}
\caption{Summary of indexes.}
\label{table:indexes}
\normalsize
\end{table}

\subsubsection{Path Indexes}
Let $\lambda\in(0,1]^{|E|}$, $t\in \NN^{|E|}$, and $n\in \NN$. The first path index, denoted by $b_p(n,\lambda,t)$ for path $p\in {\cal P}$, is motivated by the index defined in \cite{combes2015stochastic}. $b_p(n,\lambda,t)$ is defined as the infimum of the following optimization problem:
\begin{align*}
\inf_{u\in (0,1]^{|E|}} & \;\; p^\top u^{-1}\\
\hbox{subject to:} & \;\; \sum_{i\in p} t_i\kl(\lambda_i, u_i)\le f_1(n),\\
& \;\; u_i\ge \lambda_i, \;\; \forall i\in E,
\end{align*}
where $f_1(n)=\log(n)+4H\log(\log(n))$, and for all $a,b\in [0,1]$, $\kl(a,b)$ is the KL-divergence number between two Bernoulli distributions with respective means $a$ and $b$, i.e., $\kl(a,b)=a\log(a/b)+(1-a)\log((1-a)/(1-b))$.

The second index is denoted by $c_p(n,\lambda,t)$ and defined for path $p\in \Pcal$ as:
\begin{align*}
c_p(n,\lambda,t) = p^\top \lambda^{-1} - \sqrt{ \sum_{i\in p} \frac{2f_1(n)}{t_i\lambda_i^3}}.
\end{align*}

%, or for short $b_p(n)$ or for short $c_p(n)$
The next theorem provides generic properties of the two indexes $b_p$ and $c_p$.
\begin{theorem}
\label{thm:geocombucb_properties}
%(i) For all $n\ge 1$, $i\in E$, and $\lambda\in (0,1]$, we have $\omega_i(n,\tau) \geq \gamma_i(n,\tau)$.
(i) For all $n\ge 1$, $p\in {\cal P}$, $\lambda\in (0,1]^{|E|}$, and $t\in \mathbb{N}^{|E|}$, we have $b_p(n,\lambda,t) \geq c_p(n,\lambda,t)$.

(ii) There exists a constant $K_H > 0$ depending on $H$ only such that, for all $p\in {\cal P}$ and $n\ge 2$:
% $$
% \PP[b_p(n,\hat\theta(n)) \leq p^\top \theta] \leq K_H n^{-1} % (\log(n))^{-2}.
% $$
$$
\PP[b_p(n,\hat\theta(n),t(n)) \geq p^\top \theta] \leq K_H n^{-1} (\log(n))^{-2}.
$$
\end{theorem}
\begin{corollary} \label{corr:index_ge_Dstar}
We have:
\begin{align*}
\sum_{n \geq 1} \PP[b_{p^\star}&(n,\hat\theta(n),t(n)) \geq\ {p^\star}^\top \theta^{-1} ]  \\
&\leq 1+K_H \sum_{n \geq 2}  n^{-1} (\log(n))^{-2} < \infty.
\end{align*}
\end{corollary}

\subsubsection{Link Index}
Our third index is a link index. For $n,t\in \NN$ and $\lambda\in (0,1]$, the index $\omega_i(n,\lambda,t)$ of link $i\in E$ is defined as:
\begin{align*}
\omega_i(n,\lambda,t) = \min \Bigl\{\frac{1}{u}:\;u\in [\lambda,1], \;\; t\kl \bigl (\lambda, u \bigr)
  \leq f_2(n) \Big\},
\end{align*}
where $f_2(n)=\log(n)+4\log(\log(n))$.

\subsection{Routing policies}

We present three routing policies, referred to as \textsc{GeoCombUCB-1}, \textsc{GeoCombUCB-2} and \textsc{KL-SR}, respectively. For the transmission on the $n$-th packet, \textsc{GeoCombUCB-1} (resp. \textsc{GeoCombUCB-2}) selects the path $p$ with the lowest index $b_p(n):=b_p(n,\hat\theta(n),t(n))$ (resp. $c_p(n):=c_p(n,\hat\theta(n),t(n))$). \textsc{KL-SR} was initially proposed in \cite{ZouACC2014} and for the transmission of the $n$-th packet, it selects the path $p(n)\in \arg\min_{p\in \Pcal} p^\top \omega(n)$, where $\omega(n)=(\omega_i(n), i\in E)$ and $\omega_i(n):=\omega_i(n,\hat\theta_i(n),t_i(n))$. The pseudo-code of \textsc{GeoCombUCB} and \textsc{KL-SR} are presented in Algorithm \ref{alg:GeoCombUCB} and Algorithm \ref{alg:KL-SR}, respectively.

\begin{algorithm}[tb]
   \caption{\textsc{GeoCombUCB}}
   \label{alg:GeoCombUCB}
\begin{algorithmic}
   %\STATE {\bf Initialization:} For $n=1,\ldots, A$, select actions in ${\cal A}$, observe the rewards, and update $\zeta(n)$.
   \vspace{1mm}
   \FOR{$n\geq 1$}
   \STATE Select path $p(n)\in \arg\min_{p\in {\cal P}} \xi_p(n)$ (ties are broken arbitrarily), where $\xi_p(n)=b_p(n)$ for \textsc{GeoCombUCB-1}, and $\xi_p(n)=c_p(n)$ for \textsc{GeoCombUCB-2}. \vspace{1mm}
   \STATE Collect feedback on links $i \in p(n)$, and update $\hat\theta_i(n)$ for $i\in p(n)$. \vspace{1mm}
   \ENDFOR
\end{algorithmic}
\end{algorithm}

\begin{algorithm}[tb]
   \caption{\textsc{KL-SR}}
   \label{alg:KL-SR}
\begin{algorithmic}
   %\STATE {\bf Initialization:} For $n=1,\ldots, A$, select actions in ${\cal A}$, observe the rewards, and update $\zeta(n)$.
   \vspace{1mm}
   \FOR{$n\geq 1$}
   \STATE Select path $p(n)\in \arg\min_{p\in {\cal P}} p^\top \omega(n)$ (ties are broken arbitrarily).
   \vspace{1mm}
   \STATE Collect feedback on links $i \in p(n)$, and update $\hat\theta_i(n)$ for $i\in p(n)$. \vspace{1mm}
   \ENDFOR
\end{algorithmic}
\end{algorithm}

In the following theorems, we provide a finite time analysis of the \textsc{GeoCombUCB} and \textsc{KL-SR} policies and show the optimality of \textsc{KL-SR} in line networks. Define $\varepsilon= (1 - 2^{-\frac 14}) \frac{\Delta_{\min}}{D^+}$.
\begin{theorem}
\label{thm:regret_geocombucb}
 For all $N \ge 1$, under policies \newline $\pi\in\{\textsc{GeoCombUCB-1}, \textsc{GeoCombUCB-2}\}$ we have:
\eqs{
R^\pi(N) \le {16 |E| \sqrt{H} f_1(N) \over \Delta_{\min} \theta_{\min}^2} + 2 D^+ \left( 2 K_H +  \sum_{i \in E} {1 \over \varepsilon^{2} \theta_{i}^2} \right).
}
Hence $R^{\pi}(N) = {\cal O}\left(\frac{|E|\sqrt{H}}{\Delta_{\min}\theta_{\min}^2} \log(N)\right)$ when $N \to \infty$.
\end{theorem}
\vspace{4mm}

\begin{theorem}
\label{thm:regret_geocombucb_2}
For all $N \ge 1$, under policy $\pi=\textsc{KL-SR}$ we have:
\eqs{
R^\pi(N) \le {360 |E| H f_2(N) \over \Delta_{\min}  \theta_{\min}^{2}} + 2 D^+ \left( 4 H + \sum_{i \in E} {1 \over \varepsilon^2 \theta_{i}^2} \right).
}
Hence $R^{\pi}(N) = {\cal O}\left(\frac{|E|H}{\Delta_{\min}\theta_{\min}^2} \log(N)\right)$ when $N \to \infty$.
\end{theorem}
\vspace{4mm}

The index $b_p$ is an extension of the KL-based index of \cite{combes2015stochastic} to the case of geometrically distributed rewards. However the proof of
Theorem \ref{thm:regret_geocombucb} is novel and uses the link  between $b_p$ and $c_p$ established in Theorem \ref{thm:geocombucb_properties}. The proof of Theorem \ref{thm:regret_geocombucb} uses some of ideas from \cite{combes2015stochastic}. The proof of Theorem \ref{thm:regret_geocombucb_2} is completely different from the regret analysis of KL-SR in \cite{ZouACC2014}; it relies on Lemma \ref{lem:index_KL_SR}, which provides a tight lower bound for the index $\omega_i$, and borrows some ideas from \cite[Theorem~5]{kveton2014tight}.

\begin{remark}
	Theorem~\ref{thm:regret_geocombucb_2} holds even when the delays on the various links are not independent as in \cite{kveton2014tight}.
\end{remark}

The proposed policies have better performance guarantees than existing routing algorithms. Indeed, as shown in %\cite[Appendix~J]{routing_paper_TechRep},
Appendix \ref{sec:CUCB_regret}, 
the best regret upper bound for the CUCB algorithm \cite{chen2013combinatorial_icml} is
$
R^{\textrm{CUCB}}(N)={\cal O}\left(\frac{|E|H}{\Delta_{\min}\theta_{\min}^3}\log(N)\right),
$
%We believe that applying the proof techniques presented in \cite{kveton2014tight} (see the proof of Theorem 5 there), one might provide a regret upper bound for CUCB scaling as {\color{blue} ${\cal O}(\frac{|E|H}{\Delta_{\min}\theta_{\min}^4}\log(N))$},
which constitutes a weaker performance guarantee than those of our routing policies. The numerical experiments presented in the next section will confirm the superiority of \textsc{GeoCombUCB} and KL-SR over CUCB. The next proposition states that \textsc{KL-SR} is asymptotically optimal in line networks.
\medskip
\begin{proposition}
\label{prop:GeoCombUCB_Line}
In line networks, the regret under $\pi=\textsc{KL-SR}$ satisfies
%\begin{align*}
$\limsup_{N \rightarrow \infty} \dfrac{R^\pi(N)}{\log(N)} \leq c_2(\theta)$.
%\end{align*}
Hence, $R^{\pi}(N)={\cal O}\left(\frac{|E|-H}{\Delta_{\min}\theta_{\min}^2} \log(N)\right)$ when $N \to \infty$.
\end{proposition}
\vspace{3mm}
\begin{remark}
When the link parameters smoothly evolve over time, we can modify the proposed routing policies so that routing decisions are based on past choices and observations over a sliding window consisting of a fixed number of packets, as considered in \cite{garivier2008upper} and \cite{combes2014unimodal_techreport}.
\end{remark}

\subsection{Implementation}

Next we discuss the implementation of our routing policies, and give simple methods to compute $b_p(n,\lambda,t)$,  $c_p(n,\lambda,t)$, $\omega_i(n,\lambda,t)$ given $p,i,n,\lambda$ and $t$. The path index $c_p$ is explicit and easy to compute. The link index $\omega_i$ is also  straightforward as it amounts to finding the roots of a strictly convex and increasing function in one variable (note that $v\mapsto\kl(u,v)$ is strictly convex and increasing for $v\ge u$). Hence, the index $\omega_i$ can be computed by a simple line search. The path index $b_p(n,\lambda,t)$ can also be computed using simple line search, as shown below.

Define $I_p(\lambda) = \{ i\in p:  \lambda_i \neq 1\}$, and for $\gamma > 0$, define:
\als{
F(\gamma,\lambda,n,t) &= \sum_{i \in I_p(\lambda)} t_i \kl( \lambda_i(n), g(\gamma,\lambda_i,t_i)), \text{     with } \\
g(\gamma,\lambda_i,t_i)&= \frac{1}{2 \gamma t_i}\Bigl(\gamma \lambda_i t_i -1 + \sqrt{ (1 -  \gamma \lambda_i t_i)^2
+ 4 \gamma t_i}\Big).
}
\begin{proposition}
\label{prop:index_computation}
(i)  $\gamma \mapsto F(\gamma,\lambda,n,t)$ is strictly increasing, and $F(\RR^+,\lambda,n,t) = \RR^+$.
(ii) If $I_p(\lambda) = \emptyset$, $b_p(n,\lambda,t) = \sum_{i\in E} p_i$. Otherwise, let $\gamma^\star$ is the unique solution to $F(\gamma,\lambda,n,t) = f_1(n)$. Then,
$$
b_p(n,\lambda,t) = \sum_{i\in E} p_i - |I_p(\lambda)| + \sum_{i \in I_p(\lambda)} g(\gamma^\star,\lambda_i,t_i).
$$
\end{proposition}
%\vspace{3mm}
As stated in Proposition \ref{prop:index_computation}, {\color{black}proven in Appendix \ref{sec:index_computation}}, %\cite[Appendix~I]{routing_paper_TechRep}}, 
$\gamma^\star$ can be computed efficiently by a simple line search, and $b_p$ is easily deduced. We thus have efficient methods to compute the three indexes. To implement our policies, we then need to find in each round, the path maximizing the index (or the sum of link indexes along the path for KL-SR). \textsc{KL-SR} can be implemented (in a distributed fashion) using the Bellman-Ford algorithm, and its complexity is ${\cal O}(|V||E|)$ in each round.
\textsc{GeoCombUCB-1} and \textsc{GeoCombUCB-2} are more computationally involved than \textsc{KL-SR} and have complexity ${\cal O}(|{\cal P}|)$ in each round.

\subsection{Numerical Experiments}
\begin{figure*}[!th]
\begin{center}
\subfigure[A grid network]{
\includegraphics[width=0.32\columnwidth]{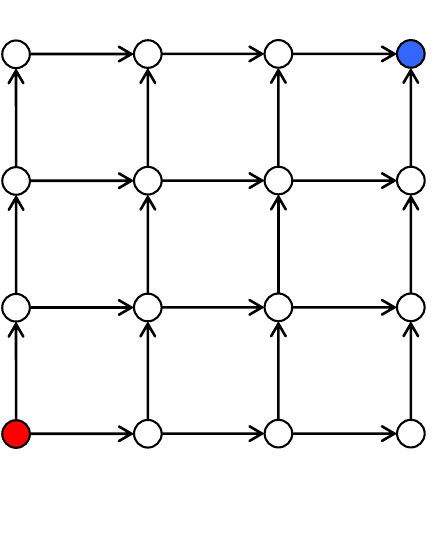}}\hspace{-2mm}
%\subfigure[$ \theta_{\min}=0.30,\; \Delta_{\min}=0.15 $]{
%\includegraphics[width=0.75\columnwidth]{205.pdf}}
\subfigure[$\theta_{\min} = 0.18,\; \Delta_{\min} = 0.34$]{
\includegraphics[width=0.85\columnwidth]{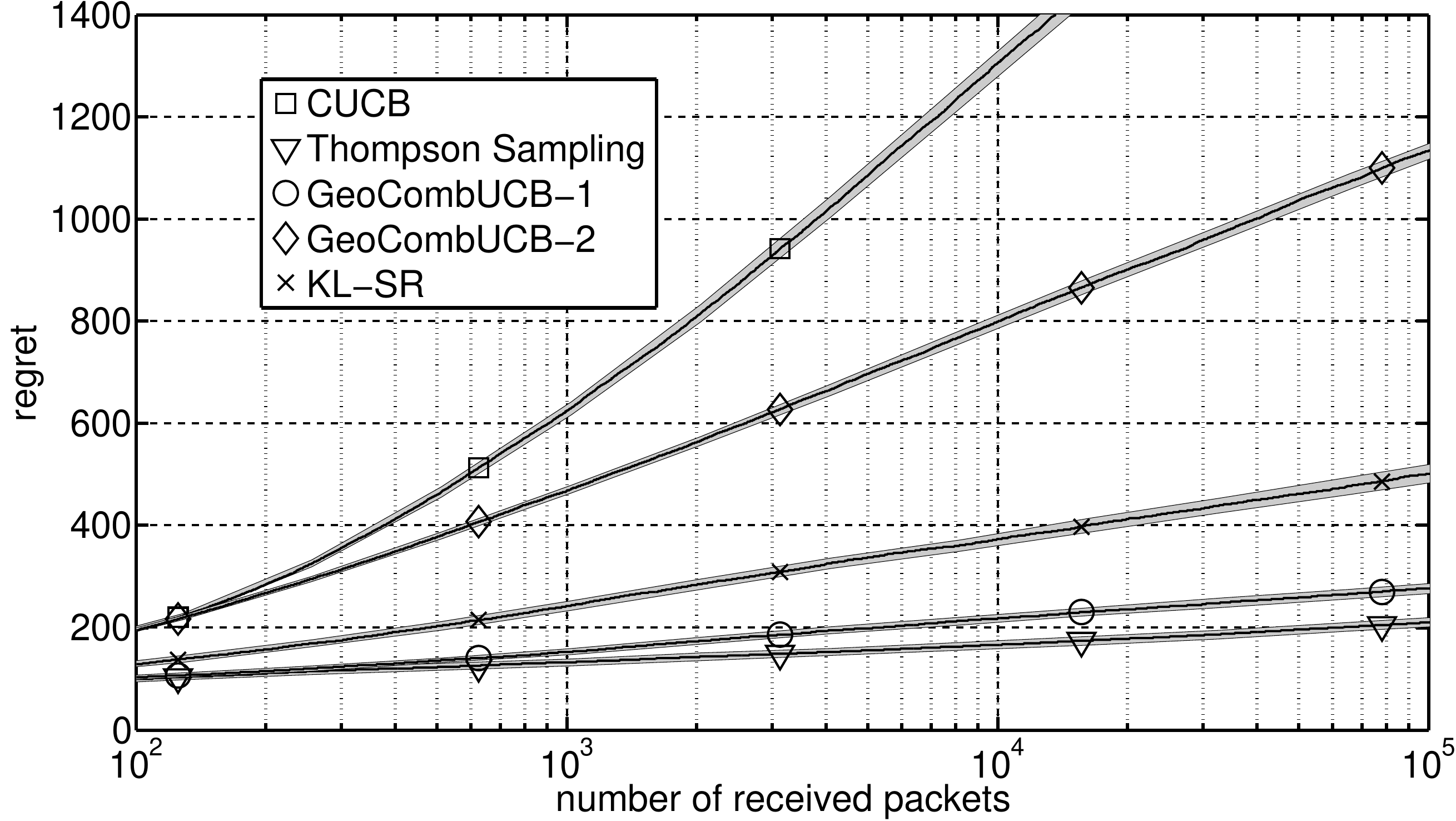}}\hspace{-2mm}
\subfigure[$\theta_{\min}=0.1,\; \Delta_{\min}=0.08$]{
\includegraphics[width=0.85\columnwidth]{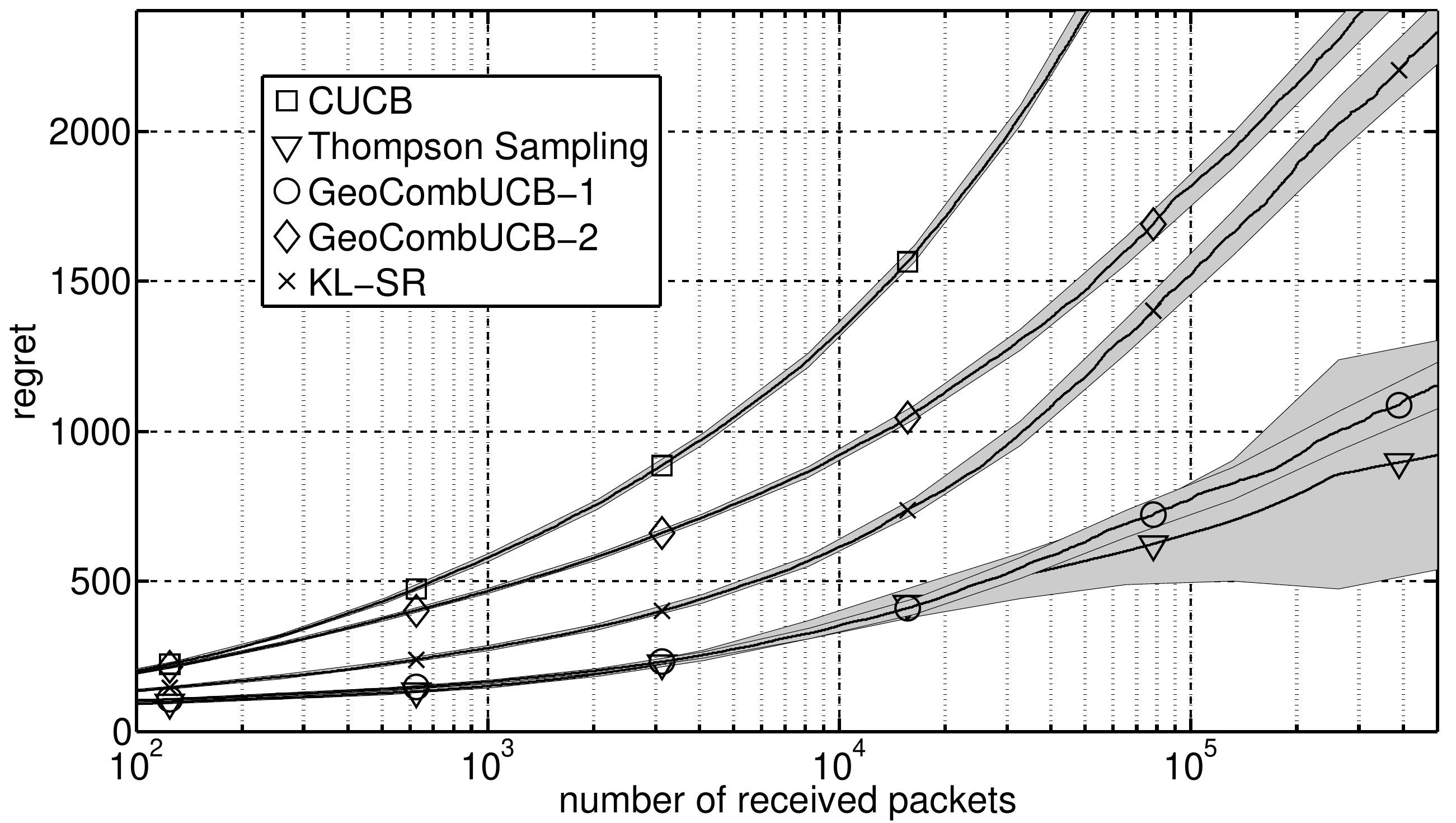}}
\end{center}
\caption{Network topology, and regret versus number of received packets.} % under different randomly generated link success probabilities.}
\label{fig:Sim_result}
\end{figure*}

In this section, we conduct numerical experiments to compare the performance of the proposed source-routing policies to that of the CUCB algorithm~\cite{chen2013combinatorial_icml} {\color{black} and TS} applied to our online routing problem. The CUCB algorithm is an index policy in $\Pi_2$ (the set of source-routing policies with semi-bandit feedback), and selects path $p(n)$ for the transmission of the $n$-th packet:
$$
p(n)\in \arg\min_{p\in \Pcal} \sum_{i\in p}\frac{1}{\hat{\theta}_i(n) + \sqrt{1.5 \log(n) / t_i(n)}}.
$$
We consider a grid network whose topology is depicted in Figure~\ref{fig:Sim_result}(a), where the node in red (resp.~blue) is the source (resp.~the destination). In this network, there are ${6\choose 3}=20$ possible paths from the source to the destination.
{\color{black} Let us compare these algorithms in terms of their per-packet complexity. The complexity of \textsc{GeoCombUCB-1} and \textsc{GeoCombUCB-2} is ${\cal O}(|{\cal P}|)$, whereas that of KL-SR, CUCB, and TS is ${\cal O}(|V||E|)$.}

In Figures~\ref{fig:Sim_result}(b)-(c), we plot the regret against the number of the packets $N$ under the various routing policies, and for two sets of link parameters $\theta$. For each set, we choose a value of $\theta_{\min}$ and generate the values of $\theta_i$ independently, uniformly at random in $[\theta_{\min},1]$. The results are averaged over $100$ independent runs, and the $95\%$ confidence intervals are shown using the grey area around curves. The three proposed policies outperform CUCB, and \textsc{GeoCombUCB-1} attains the smallest regret amongst the proposed policies.
The comparison between \textsc{GeoCombUCB-2} and \textsc{KL-SR} is more subtle and depends on the link parameters: while in Figure~\ref{fig:Sim_result}(b) KL-SR significantly outperforms \textsc{GeoCombUCB-2}, they attain regrets growing similarly for the link parameter of  Figure~\ref{fig:Sim_result}(c). Yet there are some parameters for which KL-SR is significantly outperformed by \textsc{GeoCombUCB-2}. \textsc{KL-SR} seems to perform better than \textsc{GeoCombUCB-2} in scenarios where $\Delta_{\min}$ is large. TS performs slightly better than \textsc{GeoCombUCB-1} on average. Its regret, however may not be well concentrated around the mean for some link parameters, as in Figure~\ref{fig:Sim_result}(c). Furthermore, the regret analysis of TS for shortest-path routing with general topologies is an open problem. % We refer to \cite{routing_paper_TechRep} for other numerical experiments.

\subsection{A distributed hop-by-hop routing policy}\label{app:hop}

%Motivated by this observation, we propose, in Appendix \ref{app:hop}, \textsc{KL-HHR}, a distributed routing policy which is a hop-by-hop version of \textsc{KL-SR} algorithm and hence that belongs to the set of policies $\Pi_3$.
Motivated by the Bellman-Ford implementation of \textsc{KL-SR} algorithm, we propose \textsc{KL-HHR}, a distributed routing policy which is a hop-by-hop version of \textsc{KL-SR} algorithm and hence belongs to the set of policies $\Pi_3$.
We first introduce the necessary notations.
For any node $v\in V$, we let ${\cal P}_v$ denote the set of loop-free paths from node $v$ to the destination. For any time slot $\tau$, we denote by $n(\tau)$ the packet number that is about to be sent or already in the network. For any edge $i$, let $\tilde\theta_i(\tau)$ be the empirical success rate of edge $i$ \emph{up to time slot} $\tau$, that is $\tilde\theta_i(\tau)=s_i(n(\tau))/t'_i(\tau)$, where $t'_i(\tau)$ denotes the total number of transmission attempts on link $i$ up to time slot $\tau$. Moreover, with slight abuse of notation, we denote the index of link $i$ at time $\tau$ by $\omega_i(\tau,\tilde\theta_i(\tau))$. Note that by definition $t'_i(\tau)\ge t_i(n)$ and $\tilde\theta_i(\tau)$ is a more accurate estimate of $\theta_i$ than $\hat\theta_i(n(\tau))$. %in lieu of $\hat\theta_i(n)$ gives better performance as the packet might avoid some bad links with overestimated success probability.
We define $J_v(\tau)$ as the minimum {\it cumulative index} from node $v$ to the destination:
$$
J_v(\tau) = \min_{p\in {\cal P}_v} \sum_{i\in p} \omega_i(\tau,\tilde\theta_i(\tau)).
$$
We note that $J_v(\tau)$ can be computed using Bellman-Ford algorithm. \textsc{KL-HHR} works based on the following idea: at time $\tau$ if the current packet is at node $v$, it will be sent to node $v'$ with $(v,v')\in E$ such that $\omega_{(v,v')}(\tau,\tilde\theta_v(\tau)) + J_{v'}(\tau)$ is minimal over all outgoing edges of node $v$. The pseudo-code of \textsc{KL-HHR} is given in Algorithm~\ref{alg:HopByHop}.

\begin{algorithm}[th]
\caption{\textsc{KL-HHR} for node $v$}
\begin{algorithmic}
  \label{alg:HopByHop}
    \FOR{$\tau\geq 1$}
      \STATE Select link $(v,v')\in E$, where
       \begin{align*}
       v' \in \arg \min_{w\in V: (v,w)\in E} \left(\omega_{(v,w)}(\tau,\tilde\theta_v(\tau)) + J_{w}(\tau) \right).
       \end{align*}
	\STATE Update index of the link $(v,v')$.
    \ENDFOR
\end{algorithmic}
\end{algorithm}

\begin{figure*}[!th]
\begin{center}
\subfigure[Topology]{
\includegraphics[width=0.33\columnwidth]{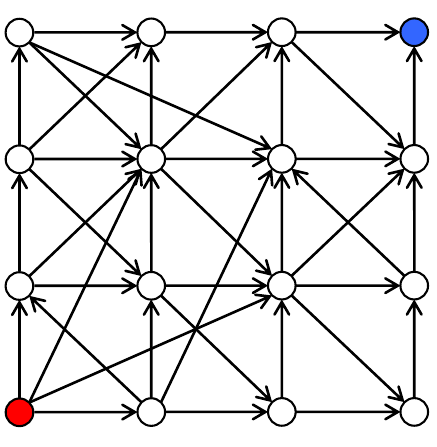}}
\subfigure[$\theta_{\min} = 0.014$]{
\includegraphics[width=0.81\columnwidth]{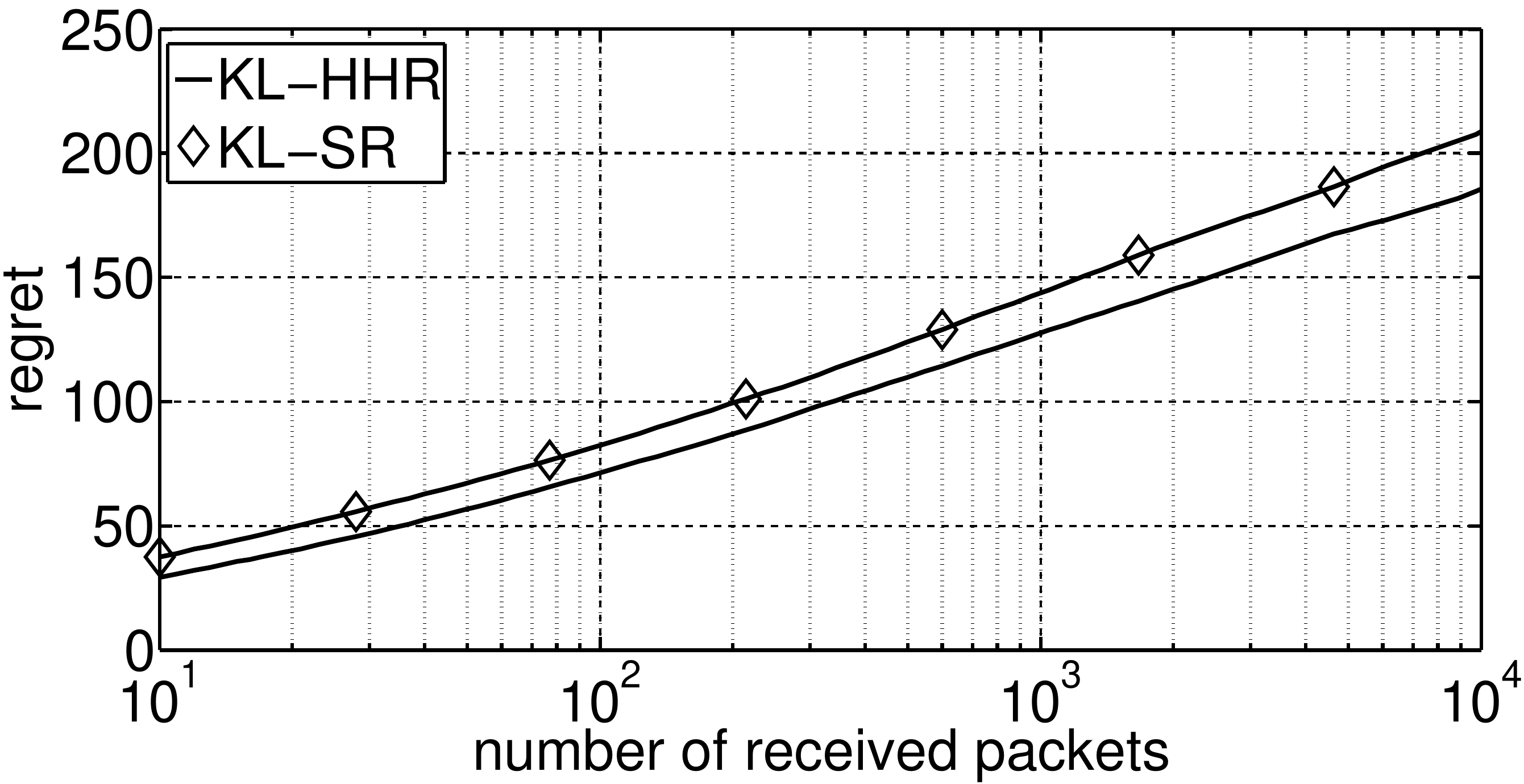}}
\subfigure[$\theta_{\min}=0.0056$]{
\includegraphics[width=0.81\columnwidth]{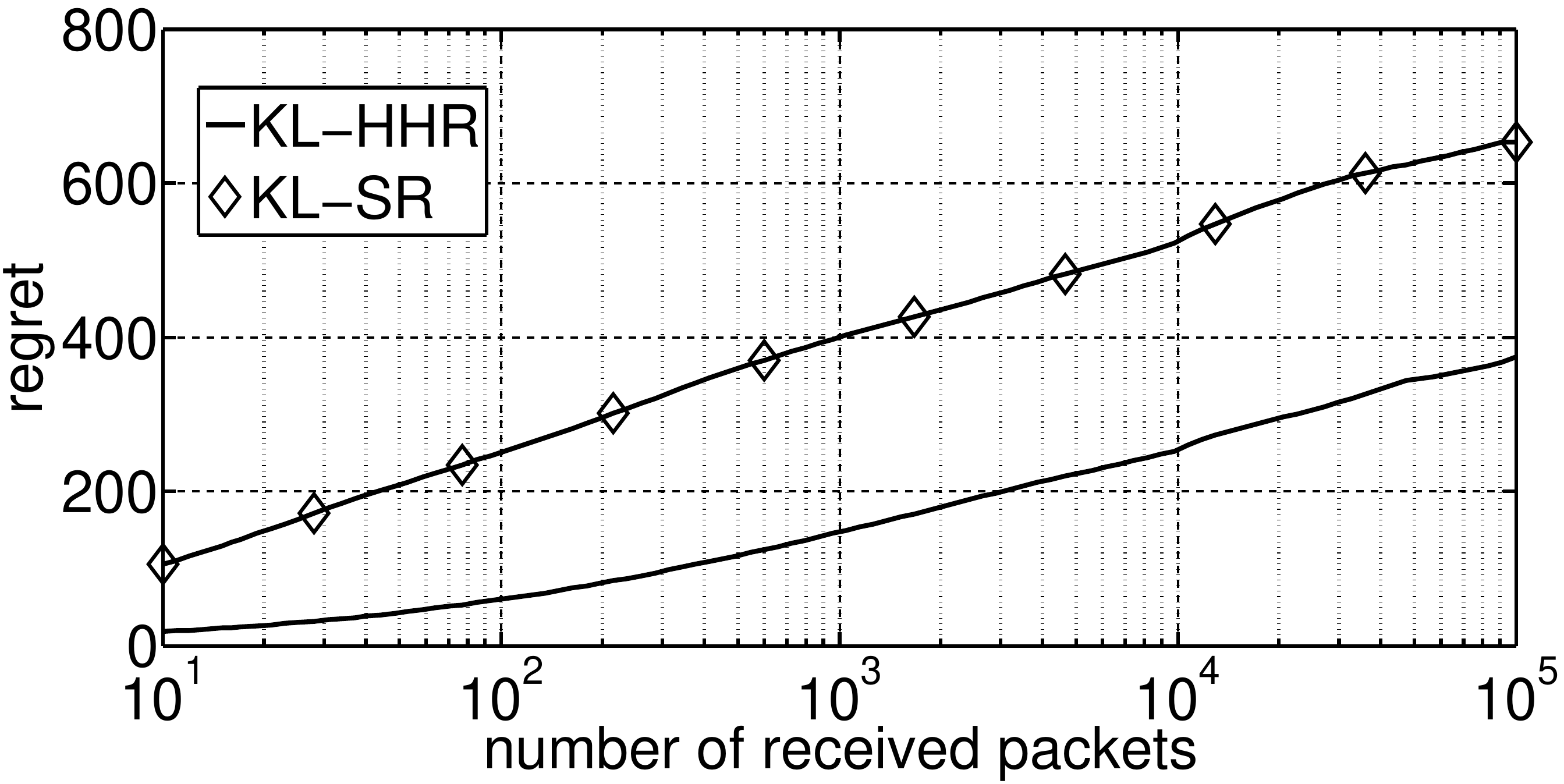}}
\end{center}
\caption{Network topology, and regret versus number of received packets.} % under different randomly generated link success probabilities.}
\label{fig:Sim_result_HbH}
\end{figure*}

We compare the performance of KL-HHR and KL-SR through numerical experiments. We consider a grid network whose topology is depicted in Figure~\ref{fig:Sim_result_HbH}(a), in which there are $40$ links and $413$ possible paths from the source (in red) to the destination (in blue).
%where the node in red (resp.~blue) is the source (resp.~the destination). In this network, there are ${6\choose 3}=20$ possible paths from the source to the destination.
Figures \ref{fig:Sim_result_HbH}(b)-(c) display the regret against the number of the packets $N$ under KL-SR and KL-HHR for two sets of link parameters $\theta$. The values of $\theta_i$ are generated similarly to the previous experiments and the results are averaged over $100$ independent runs.
As expected, KL-HHR outperforms KL-SR in both scenarios, and the difference is significant when $\theta_{\min}$ is small. The reason is that KL-HHR can change routing decisions dynamically at intermediate nodes, and does not waste transmissions on bad links when they are discovered. It is noted, however that, irrespective of the value of $\theta_{\min}$, the regret of both KL-HHR and KL-SR grow similarly when the number of received packets grows large.

The regret analysis of \textsc{KL-HHR} is beyond the scope of this paper, and is left for future work.

%The regret analysis of \textsc{KL-HHR} is beyond the scope of this paper, and is left for future work. Due to space limit, we refer to \cite{routing_paper_TechRep} for the numerical experiments.

\section{Conclusions and Future Work}
\label{sec:conclusion}

We have studied online shortest path routing problems in networks with stochastic link delays. The distributions of these delays are initially unknown, and have to be estimated by actual packet transmissions. Three types of routing policies are analyzed: source-routing with semi-bandit feedback, source-routing with bandit feedback, and hop-by-hop routing. Tight asymptotic lower bounds on the regret for the three types of policies are derived. By comparing these bounds, we observed that semi-bandit feedback significantly improves performance while hop-by-hop decisions do not. Finally, we proposed several simple routing policies for semi-bandit feedback that outperform alternatives from the literature both in theory and in numerical experiments.
As future work, we plan to propose practical algorithms with provable performance bounds for hop-by-hop routing and source-routing with bandit feedback.  Furthermore, we would like to study the effect of delayed feedback on the performance as studied in, e.g., \cite{joulani2013online}.
%Lastly, we mention that a promising approach to design algorithms with smaller regret could be to incorporate empirical variance estimation into algorithm, as in UCB-V \cite{audibert2009exploration}. This idea is motivated by the observation of \cite{neufeld2014adaptive}, where it is shown, through a systematic parameter study, that in some setups KL-UCB is often outperformed by Thompson Sampling and UCB-V.

\bibliography{MabProblem}

\begin{thebibliography}{10}
\providecommand{\url}[1]{#1}
\csname url@samestyle\endcsname
\providecommand{\newblock}{\relax}
\providecommand{\bibinfo}[2]{#2}
\providecommand{\BIBentrySTDinterwordspacing}{\spaceskip=0pt\relax}
\providecommand{\BIBentryALTinterwordstretchfactor}{4}
\providecommand{\BIBentryALTinterwordspacing}{\spaceskip=\fontdimen2\font plus
\BIBentryALTinterwordstretchfactor\fontdimen3\font minus
  \fontdimen4\font\relax}
\providecommand{\BIBforeignlanguage}[2]{{%
\expandafter\ifx\csname l@#1\endcsname\relax
\typeout{** WARNING: IEEEtran.bst: No hyphenation pattern has been}%
\typeout{** loaded for the language `#1'. Using the pattern for}%
\typeout{** the default language instead.}%
\else
\language=\csname l@#1\endcsname
\fi
#2}}
\providecommand{\BIBdecl}{\relax}
\BIBdecl

\bibitem{ZouACC2014}
Z.~Zou, A.~Proutiere, and M.~Johansson, ``Online shortest path routing: The
  value of information,'' in \emph{Proceedings of American Control Conference
  (ACC)}, Jun. 2014.

\bibitem{awerbuch2004adaptive}
B.~Awerbuch and R.~D. Kleinberg, ``Adaptive routing with end-to-end feedback:
  Distributed learning and geometric approaches,'' in \emph{Proceedings of the
  36th Annual ACM Symposium on Theory of Computing (STOC)}, 2004, pp. 45--53.

\bibitem{gyorgy2006adaptive}
A.~Gy{\"o}rgy and G.~Ottucs{\'a}k, ``Adaptive routing using expert advice,''
  \emph{The Computer Journal}, vol.~49, no.~2, pp. 180--189, 2006.

\bibitem{Gyoergy2007}
A.~Gy{\"o}rgy, T.~Linder, G.~Lugosi, and G.~Ottucs{\'a}k, ``The on-line
  shortest path problem under partial monitoring,'' \emph{Journal of Machine
  Learning Research}, vol.~8, pp. 2369--2403, 2007.

\bibitem{He2013}
T.~He, D.~Goeckel, R.~Raghavendra, and D.~Towsley, ``Endhost-based shortest
  path routing in dynamic networks,'' in \emph{Proceedings of the 32nd IEEE
  International Conference on Computer Communications (INFOCOM)}, 2013, pp.
  2202--2210.

\bibitem{brun2016big}
O.~Brun, L.~Wang, and E.~Gelenbe, ``Big data for autonomic intercontinental
  overlays,'' \emph{IEEE Journal on Selected Areas in Communications}, vol.~34,
  no.~3, pp. 575--583, 2016.

\bibitem{Cesa-Bianchi2012}
N.~Cesa-Bianchi and G.~Lugosi, ``Combinatorial bandits,'' \emph{Journal of
  Computer and System Sciences}, vol.~78, no.~5, pp. 1404--1422, 2012.

\bibitem{Auer2002}
P.~Auer, N.~Cesa-Bianchi, and P.~Fischer, ``Finite-time analysis of the
  multiarmed bandit problem,'' \emph{Machine Learning}, vol.~47, pp. 235--256,
  2002.

\bibitem{chen2013combinatorial_icml}
W.~Chen, Y.~Wang, and Y.~Yuan, ``Combinatorial multi-armed bandit: General
  framework and applications,'' in \emph{Proceedings of the 30th International
  Conference on Machine Learning (ICML)}, 2013, pp. 151--159.

\bibitem{gopalan2014thompson}
A.~Gopalan, S.~Mannor, and Y.~Mansour, ``Thompson sampling for complex online
  problems,'' in \emph{Proceedings of the 31st International Conference on
  Machine Learning (ICML)}, 2014, pp. 100--108.

\bibitem{Robbins1952}
H.~Robbins, ``Some aspects of the sequential design of experiments,''
  \emph{Bulletin of the American Mathematical Society}, vol.~58, no.~5, pp.
  527--535, 1952.

\bibitem{Lai1985}
T.~L. Lai and H.~Robbins, ``Asymptotically efficient adaptive allocation
  rules,'' \emph{Advances in applied mathematics}, vol.~6, no.~1, pp. 4--22,
  1985.

\bibitem{Audibert2014}
J.-Y. Audibert, S.~Bubeck, and G.~Lugosi, ``Regret in online combinatorial
  optimization,'' \emph{Mathematics of Operations Research}, vol.~39, no.~1,
  pp. 31--45, 2014.

\bibitem{Bubeck2012towards}
S.~Bubeck, N.~Cesa-Bianchi, and S.~M. Kakade, ``Towards minimax policies for
  online linear optimization with bandit feedback,'' in \emph{Proceedings of
  the 25th Conference On Learning Theory (COLT)}, 2012.

\bibitem{neu2013efficient}
G.~Neu and G.~Bart{\'o}k, ``An efficient algorithm for learning with
  semi-bandit feedback,'' in \emph{Algorithmic Learning Theory (ALT)}.\hskip
  1em plus 0.5em minus 0.4em\relax Springer, 2013, pp. 234--248.

\bibitem{Gai2012}
Y.~Gai, B.~Krishnamachari, and R.~Jain, ``Combinatorial network optimization
  with unknown variables: Multi-armed bandits with linear rewards and
  individual observations,'' \emph{IEEE/ACM Transactions on Networking},
  vol.~20, no.~5, pp. 1466--1478, 2012.

\bibitem{kveton2014tight}
B.~Kveton, Z.~Wen, A.~Ashkan, and C.~Szepesvari, ``Tight regret bounds for
  stochastic combinatorial semi-bandits,'' in \emph{Proceedings of the 18th
  International Conference on Artificial Intelligence and Statistics
  (AISTATS)}, 2015.

\bibitem{combes2015stochastic}
R.~Combes, M.~S. Talebi, A.~Proutiere, and M.~Lelarge, ``Combinatorial bandits
  revisited,'' in \emph{Advances in Neural Information Processing Systems
  (NIPS)}, 2015.

\bibitem{Anantharam1987}
V.~Anantharam, P.~Varaiya, and J.~Walrand, ``{Asymptotically efficient
  allocation rules for the multiarmed bandit problem with multiple plays-Part
  I: IID rewards},'' \emph{IEEE Transactions on Automatic Control}, vol.~32,
  no.~11, pp. 968--976, 1987.

\bibitem{kveton2014matroid}
B.~Kveton, Z.~Wen, A.~Ashkan, H.~Eydgahi, and B.~Eriksson, ``Matroid bandits:
  Fast combinatorial optimization with learning,'' in \emph{Proceedings of the
  30th Conference on Uncertainty in Artificial Intelligence (UAI)}, 2014.

\bibitem{gai2010learning}
Y.~Gai, B.~Krishnamachari, and R.~Jain, ``Learning multiuser channel
  allocations in cognitive radio networks: A combinatorial multi-armed bandit
  formulation,'' in \emph{Proceedings of Symposium on New Frontiers in Dynamic
  Spectrum (DySPAN)}, 2010.

\bibitem{wen2015efficient}
Z.~Wen, B.~Kveton, and A.~Ashkan, ``Efficient learning in large-scale
  combinatorial semi-bandits,'' in \emph{Proceedings of the 32nd International
  Conference on Machine Learning (ICML)}, 2015, pp. 1113--1122.

\bibitem{thompson1933likelihood}
W.~R. Thompson, ``On the likelihood that one unknown probability exceeds
  another in view of the evidence of two samples,'' \emph{Biometrika}, vol.~25,
  no. 3/4, pp. 285--294, 1933.

\bibitem{Liu2012}
K.~Liu and Q.~Zhao, ``Adaptive shortest-path routing under unknown and
  stochastically varying link states,'' in \emph{Proceedings of the 10th
  International Symposium on Modeling and Optimization in Mobile, Ad Hoc and
  Wireless Networks (WiOpt)}, 2012, pp. 232--237.

\bibitem{tehrani2013distributed}
P.~Tehrani and Q.~Zhao, ``Distributed online learning of the shortest path
  under unknown random edge weights.'' in \emph{Proceedings of the 38th
  International Conference on Acoustics, Speech, and Signal Processing
  (ICASSP)}, 2013, pp. 3138--3142.

\bibitem{Burnetas1997}
A.~N. Burnetas and M.~N. Katehakis, ``Optimal adaptive policies for {Markov}
  decision processes,'' \emph{Mathematics of Operations Research}, vol.~22,
  no.~1, pp. 222--255, 1997.

\bibitem{Puterman2005}
M.~L. Puterman, \emph{{Markov Decision Processes: Discrete Stochastic Dynamic
  Programming}}.\hskip 1em plus 0.5em minus 0.4em\relax Wiley-Interscience,
  2005.

\bibitem{jaksch2010}
T.~Jaksch, R.~Ortner, and P.~Auer, ``Near-optimal regret bounds for
  reinforcement learning,'' \emph{The Journal of Machine Learning Research},
  vol.~99, pp. 1563--1600, 2010.

\bibitem{Filippi2010}
S.~Filippi, O.~Capp{\'e}, and A.~Garivier, ``Optimism in reinforcement learning
  and {Kullback-Leibler} divergence,'' in \emph{Proceedings of the 48th Annual
  Allerton Conference on Communication, Control, and Computing}, 2010, pp.
  115--122.

\bibitem{Graves1997}
T.~L. Graves and T.~L. Lai, ``{Asymptotically efficient adaptive choice of
  control laws in controlled {Markov} chains},'' \emph{SIAM Journal on Control
  and Optimization}, vol.~35, no.~3, pp. 715--743, 1997.

\bibitem{Sen1999}
A.~Sen and N.~Balakrishnan, ``{Convolution of geometrics and a reliability
  problem},'' \emph{Statistics \& Probability Letters}, vol.~43, no.~4, pp.
  421--426, Jul. 1999.

\bibitem{shapiro2009semi}
A.~Shapiro, ``Semi-infinite programming, duality, discretization and optimality
  conditions†,'' \emph{Optimization}, vol.~58, no.~2, pp. 133--161, 2009.

\bibitem{garivier2008upper}
A.~Garivier and E.~Moulines, ``On upper-confidence bound policies for
  non-stationary bandit problems,'' \emph{arXiv preprint arXiv:0805.3415},
  2008.

\bibitem{combes2014unimodal_techreport}
\BIBentryALTinterwordspacing
R.~Combes and A.~Proutiere, ``Unimodal bandits: Regret lower bounds and optimal
  algorithms,'' \emph{arXiv:1405.5096}, 2014.
\BIBentrySTDinterwordspacing

\bibitem{joulani2013online}
P.~Joulani, A.~Gy\"orgy, and C.~Szepesv\'ari, ``Online learning under delayed
  feedback,'' in \emph{Proceedings of the 30th International Conference on
  Machine Learning (ICML)}, 2013, pp. 1453--1461.

\bibitem{Magureanu2014}
S.~Magureanu, R.~Combes, and A.~Proutiere, ``Lipschitz bandits: Regret lower
  bounds and optimal algorithms,'' in \emph{Proceedings of the 27th Conference
  on Learning Theory (COLT)}, 2014.

\bibitem{garivier2016explore}
A.~Garivier, P.~M{\'e}nard, and G.~Stoltz, ``Explore first, exploit next: The
  true shape of regret in bandit problems,'' \emph{arXiv preprint
  arXiv:1602.07182}, 2016.

\bibitem{Garivier2011}
A.~Garivier and O.~Capp{\'e}, ``The {KL-UCB} algorithm for bounded stochastic
  bandits and beyond,'' in \emph{Proceedings of the 24th Conference On Learning
  Theory (COLT)}, 2011.

\end{thebibliography}

\appendices
\allowdisplaybreaks

\section{Proofs of Theorems~\ref{thm:SourceAggregate},~\ref{thm:SourceSemi-bandit} and~\ref{thm:HbH}}
\label{sec:proofLowerBounds}

To derive the asymptotic regret lower bounds, we apply the techniques used by Graves and Lai \cite{Graves1997} to investigate efficient adaptive decision rules in controlled Markov chains. We recall here their general framework. Consider a controlled Markov chain $(X_t)_{t\ge 0}$ on a countable state space ${\cal S}$ with a control set $U$. The transition probabilities given control $u\in U$ are parameterized by $\theta$ taking values in a compact metric space $\Theta$: the probability to move from state $x$ to state $y$ given the control $u$ and the parameter $\theta$ is $P(x,y;u,\theta)$. The parameter $\theta$ is not known. The decision maker is provided with a finite set of stationary control laws $G=\{g_1,\ldots,g_K\}$ where each control law $g_j$ is a mapping from ${\cal S}$ to $U$: when control law $g_j$ is applied in state $x$, the applied control is $u=g_j(x)$.
It is assumed that if the decision maker always selects the same control law $g$, the Markov chain is irreducible with respect to some maximum irreducibility measure and has stationary distribution $\pi_\theta^g$.
The reward obtained when applying control $u$ in state $x$ is denoted by $r(x,u)$, so that the expected reward achieved under control law $g$ is $\mu_\theta(g)=\sum_x r(x,g(x))\pi_\theta^g(x)$. There is an optimal control law given $\theta$ whose expected reward is denoted by $\mu_\theta^{\star}=\max_{g\in G} \mu_\theta(g)$. Now the objective of the decision maker is to sequentially apply control laws so as to maximize the expected reward up to a given time horizon $N$. The performance of the decision making scheme can be quantified through the notion of regret which compares the expected reward to that obtained by always applying the optimal control law.

\subsection{Source Routing with Bandit Feedback -- Theorem~\ref{thm:SourceAggregate}}
To prove Theorem~\ref{thm:SourceAggregate}, we construct a controlled Markov chain as follows. The state space is $\mathbb{N}$, the control set is the set of paths ${\cal P}$, and the parameter $\theta=(\theta_i,i\in E)$ defines the success rates on the various links.
The parameter $\theta$ takes value in the compact space $\Theta = [\varepsilon, 1]^{|E|}$ for $\varepsilon$ arbitrarily close to zero.
%
% There are some technical conditions that will not be satisfied if we let $\theta=0$. Namely, we may have a Markov chain with two separate communicating class.
%
The set of control laws are stationary and each of them corresponds to a given path, i.e., $G={\cal P}$. A transition in the Markov chain occurs at time epochs where a new packet is sent. The state after a transition records the end-to-end delay of the packet. Hence the transition probabilities are $P(k,l;p,\theta)=\psi_\theta^p(l)$, and do not depend on the starting state. The cost (the opposite of reward) at state $l$ is simply equal to the delay $l$. Let us fix $\theta$, and denote by $p^\star$ the corresponding optimal path. For any two sets of parameters $\theta$ and $\lambda$, we define the KL information number under path (or control law) $p$ as:
\begin{align}
\label{eq:div_bandit}
I^p(\theta,\lambda) = \sum_{l = h(p)}^{\infty} \psi_\theta^{p}(l) \log \dfrac{\psi_\theta^{p}(l)}{\psi_\lambda^{p}(l)}.
\end{align}

We have that $I^p(\theta,\lambda) =0$ if and only if the delays over path $p$ under parameters $\theta$ and $\lambda$ have the same distribution. By Lemma \ref{lem:Geo_permut}, proven at the end of this subsection, this occurs if and only if the two following sets are identical: $\{\theta_i,i\in p\}$, $\{\lambda_i,i\in p\}$. We further define $B_1(\theta)$ as the set of bad parameters $\lambda$ such that under $\lambda$, $p^\star$ is not the optimal path, and such that $\theta$ and $\lambda$ are statistically not distinguishable (they lead to the same delay distribution along path $p^\star$). Then:

\small
\begin{align*}
B_1(\theta)=\Bigl\{\lambda:  \{\lambda_i,i\in p^\star\}=\{\theta_i,i\in p^\star\},\; \min_{p\in {\cal P}} D_\lambda(p)  < D_\lambda(p^\star) \Big\}.
\end{align*}
\normalsize

By \cite[Theorem~1]{Graves1997}, we conclude that the delay regret scales at least as $c_1(\theta)\log(N)$ where
\begin{align*}
  \nonumber c_1(\theta)
   = \inf \Bigl\{
  \sum_{p \in \Pcal} x_{p} \Delta_p: x\ge 0, \; \inf_{\lambda \in B_1(\theta)}
  \sum_{p \neq p^\star} x_{p} I^p(\theta,\lambda) \geq 1 \Big\},
\end{align*}
where $I^p(\theta,\lambda)$ is given in (\ref{eq:div_bandit}).
\ep

%\begin{align*}
%  \nonumber c_1(\theta)
%   = & \inf \Biggl\{
%  \sum_{p \in \Pcal} x_{p} \Delta_p: x\ge 0,\\
%  & \hspace{8mm} \inf_{\lambda \in B_1(\theta)}
%  \sum_{p \neq p^\star} x_{p}
%  \sum_{k = h(p)}^{\infty}
%  \psi_\theta^{p}(k) \log \dfrac{\psi_\theta^{p}(k)}{\psi_\lambda^{p}(k)}
%  \geq 1 \Biggr\}.
%\end{align*}

%The above derivation does not involve a specific form of link delay distribution, and hence is valid for any parametric delay distribution by an appropriately defined delay distribution~$\psi_\theta^{p}(\cdot)$.

\medskip
\begin{lemma} \label{lem:Geo_permut}
Consider $(X_i)_i$ independent with $X_i \sim \mathrm{Geo}(\theta_i)$ and $0 < \theta_i \leq 1$. Consider $(Y_i)_i$ independent with $Y_i \sim \mathrm{Geo}(\lambda_i)$ and $0 < \lambda_i \leq 1$. Define $\overline{X} = \sum_i X_i$ and $\overline{Y} = \sum_i Y_i$. Then {\color{black}$\overline X \eqdist \overline Y$} if and only if $(\theta_i)_i = (\lambda_i)_i$ up to a permutation\footnote{The symbol $\eqdist$ denotes equality in distribution.}.
\end{lemma}
\medskip

\bp
If $(\theta_i)_i = (\lambda_i)_i$, up to a permutation then $X \eqdist Y$ by inspection. Assume that $X \eqdist Y$. Define $z_m = \min_i( \min( 1/(1 - \theta_i) , 1/(1 - \lambda_i) )$. For all $z$ such that  $|z| < z_m$ we have $\EE[z^{\overline{X}}] =  \EE[z^{\overline{Y}}]$ so that
$$ \prod_{i} \frac{\theta_i}{1 - (1 - \theta_i)z} = \prod_{i} \frac{\lambda_i}{1 - (1 - \lambda_i)z}.$$
Hence:
$$ P_{X}(z) := \prod_{i} \theta_i (1 - (1 - \lambda_i)z) =  \prod_{i} \lambda_i (1 - (1 - \theta_i)z) := P_{Y}(z).$$
Both $P_{X}(z)$ and $P_{X}(z)$ are polynomials and are equal on an open set. So they are equal everywhere, and the sets of their roots are equal $\{  1/(1 - \theta_i) , i   \} = \{  1/(1 - \lambda_i) , i   \}$. So $(\theta_i)_i = (\lambda_i)_i$ up to a permutation as announced.
\ep

\subsection{Source Routing with Semi-bandit Feedback -- Theorem~\ref{thm:SourceSemi-bandit}}

The proof of Theorem~\ref{thm:SourceSemi-bandit} is similar to that of Theorem~\ref{thm:SourceAggregate}, except that here we have to account for the fact that the source gets feedback on per-link basis. To this end, we construct a Markov chain that records the delay on each link of a path. The state space is $\mathbb{N}^{|E|}$. Transitions occur when a new packet is sent from the source, and the corresponding state records the observed delays on each link of the chosen path, and the components of the state corresponding to links not involved in the path are set equal to 0. For example, the state $(0,1,4,0,7)$ indicates that the path consisting of links 2, 3, and 5 has been used, and that the per-links delays are 1, 4, and 7, respectively. The cost of a given state is equal to the sum of its components (total delay). Now assume that path $p=(i_1,\ldots,i_{h(p)})$ is used to send a packet, then the transition probability to a state whose $i_k$-th component is equal to $d_k$, $k=1,\ldots,h(p)$ (the other components are 0) is $\prod_{k=1}^{h(p)} q_\theta(i_k,d_k)$, where $q_\theta(i,m)=\theta_i(1-\theta_i)^{m-1}$ for any link $i$ and any delay $m$.
Now the KL information number of $(\theta,\lambda)$ under path $p$ is given by
\begin{align}
\label{eq:div_semibandit}
I^{p}(\theta,\lambda) = \sum_{i \in p}  \klg(\theta_i,\lambda_i),
\end{align}
since KL divergence is additive for independent random variables.
%where $\klg(\theta_i,\lambda_i)$ is the KL divergence between two geometric distributions parameterized by $\theta_i$ and $\lambda_i$.
Hence, under semi-bandit feedback, we have $I^{p}(\theta,\lambda)=0$ if and only if $\theta_i=\lambda_i$ for all $i\in p$. The set $B_2(\theta)$ of bad parameters is defined as:
$$
B_2(\theta) = \bigl\{ \lambda: \lambda_i = \theta_i \; \forall i \in p^\star, \min_{p\in {\cal P}} D_\lambda(p) < D_\lambda(p^\star)\bigr\}.
$$
Applying \cite[Theorem~1]{Graves1997} gives:
\begin{align*}
c_2(\theta) = \inf \Bigl\{
\sum_{p\in \Pcal} x_{p}\Delta_p: x\ge 0, \;
\inf_{\lambda \in B_2(\theta)} \sum_{p\neq p^\star} x_{p} I^p(\theta,\lambda) \geq 1 \Big\},
\end{align*}
where $I^p(\theta,\lambda)$ is given in (\ref{eq:div_semibandit}).
\ep

%\begin{align*}
%c_2(\theta) = &\inf \Biggl\{
%\sum_{p\in \Pcal} x_{p}\Delta_p: x\ge 0, \\
%& \hspace{8mm} \inf_{\lambda \in B_2(\theta)} \sum_{p\neq p^\star} x_{p} \sum_{i \in p}
%    \klg(\theta_i, \lambda_i) \geq 1 \Biggr\}.
%\end{align*}

% The above derivation does not involve geometric distribution on link delays except the definition of $q_\theta(.,.)$. By choosing appropriate $q_\theta(.,.)$, we can generalize the above result to any parametric link delay distribution.

\subsection{Hop-by-hop Routing -- Theorem~\ref{thm:HbH}}

This case is more involved.
We first define another notion of regret corresponding to the achieved throughput (i.e., the number of packets successfully received by the destination per unit time).
The throughput regret is introduced to ease the analysis, since computing the throughput regret is easier in the hop-by-hop case. Define $\mu_\theta(p)$ as the average throughput on path $p$ given link success rates $\theta$: $\mu_\theta(p)=1/D_\theta(p)$.
The {\it throughput regret} $S^\pi(T)$ of $\pi$ over time horizon $T$ is:
$
S^\pi(T) := T\mu_\theta(p^\star)
- \EE \left[ N^\pi(T) \right],
$
where $N^\pi(T)$ is the number of packets received up to time $T$ under policy $\pi$. Lemma~\ref{lem:delayEqThroughput}, stated at the end of the proof, provides the relation between asymptotic bound on $R^\pi(N)$ and $S^\pi(N)$.

Now we are ready to prove Theorem~\ref{thm:HbH}.
We let the state of the Markov chain be the packet location. The action is the selected outgoing link. The transitions between two states take one time slot -- the time to make a transmission attempt. Hence, the transition probability between state $x$ and $y$ with the action of using link $i$ is denoted by (where $y \neq x$) $P_{\theta}^i (x,y)=\theta_i$  if link $i$ connects node $x$ and $y$ and is zero otherwise.
%\begin{align*}
%  P_{\theta}^i (x,y) =
%  \begin{cases}
%    \theta_i & \quad \text{if link $i$ connects node $x$ and $y$}; \\
%    0 & \quad \text{otherwise}.
%  \end{cases}
%\end{align*}
On the other hand, the probability of staying at the same state is the transmission failure probability on link $i$ if link $i$ is an outgoing link, that is $P_{\theta}^i (x,x)=1 - \theta_i$ if link $i$ is an outgoing link, and is zero otherwise.
%\begin{align*}
%  P_{\theta}^i (x,x) =
%  \begin{cases}
%    1 - \theta_i & \quad \text{if link $i$ is an outgoing link}; \\
%    1 & \quad \text{otherwise}.
%  \end{cases}
%\end{align*}

We assume that the packet is injected at the source immediately after the previous packet is successfully delivered, and we are interested in counting the number of successfully delivered packets. In order not to count the extra time slot we will spend at the destination, we use a single Markov chain state to represent both the source and the destination.

\begin{figure}[!th]
  \centerline{\includegraphics[width=1\hsize]{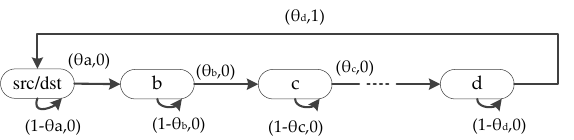}}
  \caption{A Markov chain example under a control law $p$ where the values in the parenthesis respectively denote the transition probability and the reward.}
  \label{fig:MarkovChainHopByHop}
\end{figure}

We give a reward of 1 whenever the packet is successfully delivered to the destination. Let $r(x,y,i)$ be the immediate reward after the transition from node $x$ to node $y$ under the action $i$, i.e., $r(x,y,i)=1$ if $y$ is the destination node and is zero otherwise (see Figure~\ref{fig:MarkovChainHopByHop} for an example).
%\begin{align*}
%  r(x,y,i) =
%  \begin{cases}
%    1 & \quad \text{if $y$ is the destination node}; \\
%    0 & \quad \text{otherwise}.
%  \end{cases}
%\end{align*}
Hence $r(x,i)$ (i.e., the reward at state $x$ with action $i$) is
\begin{align*}
%  r(x,i)  = \sum_{y} P_{\theta}^i (x,y) r(x,y,i)=\theta_i\bI\{\text{if link $i$ connects node $x$ to destination}\}.
r(x,i)=
\begin{cases}
  \theta_i & \; \text{if link $i$ connects node $x$ and the destination}; \\
    0 & \; \text{otherwise}.
  \end{cases}
\end{align*}
% Strictly speaking, in Graves and Lai's framework, $r(x,i)$ is a known function. However, it does not matter here since this random reward is the same as the transition probability.

The stationary control law prescribes the action at each state, i.e., the outgoing link at each node.
A stationary control law of this Markov chain is then a path $p$ in the network, and we assign arbitrary actions to the nodes that are not on the path $p$.
The maximal irreducibility measure is then to assign measure zero to the nodes that are not on the path $p$, and a counting measure to the nodes on the path $p$.
The Markov chain is irreducible with respect to this maximal irreducibility measure, and the stationary distribution of the Markov chain under path $p$ is,
\begin{align*}
  \pi^p_\theta(x) =\dfrac{\frac{1}{\theta_{p(x)}}}{\sum_{i \in p}\frac{1}{\theta_i}}\bI\{\text{if node $x$ is on the path $p$}\},
  %\begin{cases}
%   \dfrac{\frac{1}{\theta_{p(x)}}}{\sum_{i \in p}\frac{1}{\theta_i}},
%    & \text{if node $x$ is on the path $p$};  \\
%   0, & \text{otherwise},
%  \end{cases}
\end{align*}
where $p(x)$ denotes the link we choose at node $x$. The long-run average reward of the Markov chain under control law $p$ is
$
  \sum_{x} \pi^p_\theta(x) r(x,p(x)) =
  \dfrac{1}{\sum_{i \in p} \frac{1}{\theta_i}} = \mu_\theta(p).
$
The optimal control law is then $p^\star$ with long run average reward $\mu_\theta(p^\star)$.

The throughput regret of a policy $\pi\in \Pi_3$ for this controlled Markov chain at time $T$ is
\begin{align}
  S^{\pi}(T) = T \mu_\theta (p^\star) - \EE_\theta
  [\sum_{t=1}^{T} r(x_t, \pi(t,x_t))],
  % \label{eq:throughputRegret}
\end{align}
where $x_t$ is the state at time $t$ and $\pi(t,x_t)$ is the corresponding action for state $x_t$ at time $t$. To this end, we construct a controlled Markov chain that corresponds to the hop-by-hop routing in the network. Now define $I^{p} (\theta, \lambda)$ as the KL information number for a control law $p$:
\begin{align}
  &I^{p} (\theta, \lambda)  = \sum_{x} \pi^p_\theta(x)
  \sum_{y} P_{\theta}^{p(x)}(x,y)
  \log \dfrac{P_{\theta}^{p(x)} (x,y)}{P_{\lambda}^{p(x)} (x,y)} \sk
   &=   \sum_{x} \pi^p_\theta(x)
  \Big( \theta_{p(x)} \log \dfrac{\theta_{p(x)}}{\lambda_{p(x)}}
  + (1-\theta_{p(x)}) \log \dfrac{1-\theta_{p(x)}}{1-\lambda_{p(x)}}\Big) \sk
  \label{eq:div_hbh}
  &= \mu_\theta(p) \sum_{i \in p} \dfrac{\kl(\theta_i, \lambda_i)}{\theta_i}
   = \mu_\theta(p) \sum_{i\in p} \klg(\theta_i, \lambda_i),
\end{align}
where we used  Lemma~\ref{lem:KLG_KL} in the last equality.
Since $I^p(\theta,\lambda) = 0$ if and only if $\theta_i = \lambda_i$ for all $i \in p$, the set $B_2(\theta)$ of bad parameters is:
\begin{align*}
  B_2(\theta) &= \left\{ \lambda:
   \lambda_i = \theta_i \; \forall i \in p^\star,  \;
   \max_{p \in {\cal P}} \mu_\lambda(p)  > \mu_\lambda(p^\star)
    \right\} \\
    &= \left\{ \lambda: \lambda_i = \theta_i, \forall i \in p^\star, \min_{p\in {\cal P}} D_\lambda(p)
< D_\lambda(p^\star)\right\}.
\end{align*}
Applying \cite[Theorem~1]{Graves1997}, we get:\; $\liminf_{T \rightarrow \infty}
S^{\pi}(T) / \log(T) \geq c_3'(\theta)$, where
\small
\begin{align*}
c_3'(\theta)=  \inf
\Bigl\{ \sum_{p \in \Pcal} x_{p}\Delta_p:x\ge 0; \;
 \inf_{\lambda \in B_2(\theta)}
\sum_{p \neq p^\star} x_{p} \mu_\theta(p) I^p(\theta,\lambda)
 \geq 1
\Big\},
\end{align*}
\normalsize
where $I^p(\theta,\lambda)$ is given in (\ref{eq:div_hbh}).
%{\color{blue} By Lemma \ref{lem:delayEqThroughput}}, $c_3(\theta)= c_3'(\theta)/ \mu_\theta(p^\star)$. Lastly, observe that $\mu_\theta(p^\star) - \mu_\theta(p) = \mu_\theta(p^\star) \mu_\theta(p) (D_\theta(p) - D_\theta(p^\star) ).$ It follows that $c_3'(\theta)/ \mu_\theta(p^\star)=c_2(\theta)$ and therefore $c_3(\theta)=c_2(\theta)$.
 By Lemma \ref{lem:delayEqThroughput}, $c_3(\theta)\ge  c_3'(\theta)/ \mu_\theta(p^\star)$. Lastly, observe that $\mu_\theta(p^\star) - \mu_\theta(p) = \mu_\theta(p^\star) \mu_\theta(p) (D_\theta(p) - D_\theta(p^\star) ).$ It then follows that $c_3'(\theta)/ \mu_\theta(p^\star)=c_2(\theta)$ and therefore $c_3(\theta)\ge c_2(\theta)$. On the other hand, $c_3(\theta)\le c_2(\theta)$ since $\Pi_2\subset \Pi_3$. As a result, $c_3(\theta)= c_2(\theta)$ and the proof is completed.
\ep

The following two lemmas prove useful in the proof of Theorem~\ref{thm:HbH}. Lemma~\ref{lem:KLG_KL} follows from a straightforward calculation, and relates the KL-divergence between two geometric distributions to that of corresponding Bernoulli distributions. Lemma~\ref{lem:delayEqThroughput} provides the connection between the throughput regret $S^\pi(T)$ and delay regret $R^\pi(N)$ and its proof is provided in the next section.
\medskip

\begin{lemma}
\label{lem:KLG_KL}
For any $u,v \in (0,1]$, we have:
\eq{\klg(u, v)= \dfrac{\kl(u, v)}{u}. \label{lem:BerEqualGeo} }
\end{lemma}
\medskip

\bp
We have:
 \begin{align*}
 \klg(u, v) = &  \sum_{i=1}^{\infty} \bigg[
   \log \dfrac{u (1-u)^{i-1} }{v (1-v)^{i-1}} \bigg]
   u (1-u)^{i-1} \\
 =& \sum_{i=1}^{\infty}(\log \dfrac{u}{v})
    u(1-u)^{i-1} \sk
+& \sum_{i=1}^{\infty} (i-1) (\log \dfrac{1-u}{1-v})
   u (1-u)^{i-1} \\
 =& \log \dfrac{u}{v} + ( \log \dfrac{1-u}{1-v} )
    \dfrac{1-u}{u}= \dfrac{\kl(u, v)}{u}.
\end{align*}
\ep
\medskip

\begin{lemma}
\label{lem:delayEqThroughput}
For any $\pi\in \Pi_i$, $i=1,2,3$, and any $\beta>0$ we have:
\begin{align*}
\liminf_{T\to\infty}
\dfrac{ S^\pi(T) }{ \log(T) } \ge \beta \;\Longrightarrow \; \mu_\theta(p^\star)\liminf_{N\to\infty}
\dfrac{ R^\pi(N) }{ \log(N) } \ge \beta.
\end{align*}
\end{lemma}
\medskip

\section{Proof of Lemma  \ref{lem:delayEqThroughput}}

\bp
Define $\mu^\star=\mu_\theta(p^\star)$ and $r_t = \sum_{n=1}^t (D^\pi(n) - D^\star)$. Since $T \le \sum_{n=1}^{N^\pi(T) + 1} D^\pi(n)$ and $\mu^\star = {1 \over D^\star}$:
$$
T \mu^\star - N^\pi(T) \le  1 + \sum_{n=1}^{N^\pi(T) + 1} (\mu^\star D^\pi(n) - 1) =   1 + \mu^\star r_{N^\pi(T) + 1}.
$$
Since $r_t$ is a submartingale, $N^\pi(T)$ is a stopping time and $N^\pi(T) \le T$ a.s., Doob's optional stopping theorem gives:
$$
\EE(r_{N^\pi(T) + 1}) \le \EE(r_{T+1}) = R^{\pi}(T+1).
$$
Taking expectations above yields:
$$
{S^{\pi}(T) \over \log(T)} \le {1 \over \log(T)} + \mu^\star {R^{\pi}(T+1) \over \log(T)},
$$
and letting $T \to \infty$ proves the result since ${\log(T) \over \log(T+1)} \to 1$.
\ep

\section{Proof of Lemma~\ref{prop:LineNetwork_Ctheta}}

\subsection{Lower bound for $c_1(\theta)$}
Let us first decompose the set $B_1(\theta)$.
 Observe that $\min_{p\in\Pcal} D_\lambda(p) < D_\lambda(p^\star)$, implies that \emph{at least one} sub-optimal link $i$ should have a higher success probability than the link $\zeta(i)$ under the parameter $\lambda$.
Hence, we decompose $B_1(\theta)$ into sets where the link $i$ is better than the link $\zeta(i)$ under parameter $\lambda$. For any $i\notin p^\star$, define
\begin{align*}
  A_i(\theta) = & \Bigl\{ \lambda:
  \{ \lambda_{j},  j \in p^\star\}
  = \{ \theta_{j}, j \in p^\star\},
  \lambda_i > \theta_{\zeta(i)}
  \Big\}. %
\end{align*}%
Then, $B_1(\theta) = \bigcup_{i \neq \zeta(i)}  A_i(\theta)$ and Eq.~\eqref{eq:SourceRoutingAgg} reads
\begin{align*}
c_1(\theta) \; = \;\inf_{x\ge 0}& \; \; \sum_{p\in \Pcal} x_p \Delta_p
\\
\hbox{subject to:}&
\;\;
 \inf_{\lambda \in A_i(\theta)} \sum_{p \neq p^\star} x_{p} I^p( \theta,\lambda)
  \geq 1, \;\; \forall i\notin p^\star.
\end{align*}

Let $i\notin p^\star$. Consider $\vartheta^{i}$ with $\vartheta^{i}_i=\theta_{\zeta(i)}$ and $\vartheta^{i}_j=\theta_j$ for $j\neq i$. Since $\vartheta^{i}\in A_{i}(\theta)$, we have
\begin{align*}
  \inf_{\lambda \in A_{i}(\theta)}
  \sum_{p \neq p^\star} x_p
  I^p(\theta,\lambda)&\le   \sum_{p: i\in p} x_p
  I^p(\theta, \vartheta^{i}) \\
  &\le \max_{p:i\in p} I^p(\theta,\vartheta^i) \sum_{p:i\in p} x_p.
\end{align*}%
Moreover, we have that
\begin{align*}
  \sum_{p\in \Pcal} x_p\Delta_p &= \sum_{p\in \Pcal} x_p \sum_{i \in p}
   \left(\dfrac{1}{\theta_i} - \dfrac{1}{\theta_{\zeta(i)}}\right) \sk
   &= \sum_{i\notin p^\star}
   \left(\dfrac{1}{\theta_i} - \dfrac{1}{\theta_{\zeta(i)}}\right) \sum_{p: i \in p } x_p.
\end{align*}
Putting these together yields
\begin{align*}
c_1(\theta)\; \ge \; \inf_{x\ge 0}& \;\;  \sum_{i\notin p^\star}
   \left(\dfrac{1}{\theta_i} - \dfrac{1}{\theta_{\zeta(i)}}\right) \sum_{p: i \in p } x_p\\
\text{subject to:} &\;\; (\max_{p:i \in p} I^p(\theta, \vartheta^{i}))\sum_{p:i\in p} x_{p}  \geq 1, \;\; \forall i \notin p^\star.
\end{align*}
Introducing $z_i=\sum_{p: i \in p } x_p$ for any $i$, we rewrite the above problem as:
 \begin{align*}
c_1(\theta) \; \ge \; \inf_{z\ge 0}& \;\;  \sum_{i\notin p^\star}
   \left(\dfrac{1}{\theta_i} - \dfrac{1}{\theta_{\zeta(i)}}\right) z_i\\
\text{subject to:} & \;\; z_i  \geq (\max_{p:i \in p} I^p(\theta, \vartheta^{i}))^{-1}, \;\; \forall i \notin p^\star,
\end{align*}
thus giving:
\begin{align*}
c_1(\theta) \ge \sum_{i\notin p^\star}
\dfrac{ \frac{1}{\theta_i} - \frac{1}{\theta_{\zeta(i)}} }
{\max_{p:i\in p} I^p(\theta,\vartheta^i)},
\end{align*}
where $I^p(\cdot,\cdot)$ is given by (\ref{eq:div_bandit}).

\subsection{Derivation of $c_2(\theta)$}
Let us first decompose the set $B_2(\theta)$.
We argue that \mbox{$\min_{p\in\Pcal} D_\lambda(p) < D_\lambda(p^\star)$} implies that \emph{at least one} sub-optimal link $i$ should have a higher success probability than the link $\zeta(i)$ under parameter $\lambda$.

% The path $p^i$ is then better than the path $p^\star$ under parameter $\lambda$.
% Recall that $\zeta(i)$ is the optimal link on the same hop as the link $i$ under the parameter $\theta$.
We let $A_i(\theta)$ be the set where link $i$ is better than the link $\zeta(i)$ under parameter $\lambda$:
\begin{align*}
   A_i(\theta) = \bigl\{ \lambda:
   (\lambda_{j} = \theta_{j}, \; \forall j \in p^\star), \;
    \lambda_i > \theta_{\zeta(i)} \bigr\}.
\end{align*}
Hence,
  $B_2(\theta) = \bigcup_{i\notin p^\star}  A_i(\theta).$
Note $\klg(u,v) = 0$ if and only if $u=v$ and it is monotone increasing in $v$ in the range $v > u$.
Thus, for any $\lambda \in A_i(\theta)$, the infimum is obtained when $\lambda_i = \theta_{\zeta(i)}$ and $\lambda_{j} = \theta_{j} \; \forall j \neq i$, so that
\begin{align*}
  &\inf_{\lambda \in A_i(\theta)}
  \sum_{p \neq p^\star} x_{p} \sum_{i \in p}
    \klg(\theta_{i}, \lambda_{i}) \geq 1 \sk
\Longleftrightarrow\quad
&
\klg(\theta_i, \theta_{\zeta(i)})
\sum_{{p} : i \in {p}} x_{p} \geq 1.
\end{align*}
Defining $z_i = \sum_{p: i \in p } x_p$ for any $i$ and recalling that
\mbox{$
\sum_{p\in \Pcal} x_p\Delta_p
   = \sum_{i\notin p^\star}
   \left(\frac{1}{\theta_i} - \frac{1}{\theta_{\zeta(i)}}\right) \sum_{p: i \in p } x_p,
$}
we rewrite problem \eqref{eq:SourceRoutingDetail} as
\begin{align*}
  \inf_{z\ge 0} & \quad \sum_{i\notin p^\star}
   \left(\dfrac{1}{\theta_i} - \dfrac{1}{\theta_{\zeta(i)}}\right) z_i \\
   \text{subject to:} &
   \quad \klg(\theta_i, \theta_{\zeta(i)}) z_i \geq 1,
   \quad \forall i\notin p^\star,
\end{align*}
which gives
\begin{align*}
  c_2(\theta)= \sum_{i\notin p^\star}
  \dfrac{\frac{1}{\theta_i} - \frac{1}{\theta_{\zeta(i)}}}
  {\klg(\theta_i, \theta_{\zeta(i)})}
\end{align*}
and concludes the proof.
\ep

\section{Proof of Proposition \ref{prop:LineNetwork_LB_example}}
\label{sec:Line_network_LB}

\bp
Consider a problem instance with line topology in which $\theta_i=\alpha$ for all $i\notin p^\star$, and $\theta_{i}=\alpha+\alpha^2$ for all $i\in p^\star$ for some $\alpha \in (0, 0.36]$. Hence,  $\theta_i<0.5$ for all $i\in p^\star$. For any uniformly good policy $\pi\in \Pi_2\cup \Pi_3$, by Lemma \ref{prop:LineNetwork_Ctheta} we have that:

\begin{align*}
\liminf_{N\to \infty} &\frac{R^\pi(N)}{\log(N)}\ge \sum_{i\notin p^\star} \frac{1}{\klg(\theta_i,\theta_{\zeta(i)})}\Bigl(\frac{1}{\theta_i}-\frac{1}{\theta_{\zeta(i)}}\Big) \sk
 %\label{eq:LBexample_kl}
&\ge \sum_{i\notin p^\star} \frac{1}{2(\theta_{\zeta(i)}-\theta_i)} =  \sum_{i\notin p^\star} \frac{1}{2\theta_i\theta_{\zeta(i)}(\theta_i^{-1}-\theta_{\zeta(i)}^{-1})} \sk
&= \frac{|E|-H}{2\alpha(\alpha+\alpha^2)(\alpha^{-1}-(\alpha+\alpha^2)^{-1})} \sk
&= \frac{|E|-H}{2\alpha(\alpha+\alpha^2)\Delta_{\min}} \ge \frac{|E|-H}{4\alpha^2\Delta_{\min}}=\frac{|E|-H}{4\theta_{\min}^2\Delta_{\min}}, \nonumber
\end{align*}
where in the second inequality we used Lemma \ref{lem:KLG_KL} and $\kl(u,v)\le \frac{(u-v)^2}{v(1-v)} \le \frac{2(u-v)^2}{v}$ for $v\le 0.5$.
This implies that the regret of any uniformly good policy $\pi\in \Pi_2\cup \Pi_3$ for this problem instance is at least \mbox{$\Omega\Bigl(\frac{|E|-H}{\Delta_{\min}\theta_{\min}^{2}}\log(N)\Big)$}.
\ep 
\section{Proof of Theorem \ref{thm:geocombucb_properties}}
\label{sec:GeoCombUCBProperties}

We first recall two results. Lemma~\ref{lem:colt} is a concentration inequality derived in
\cite[Theorem~2]{Magureanu2014}. Lemma~\ref{lem:refined_Pinsker_Ineq}, proven in \cite[Lemma~6]{garivier2016explore}, is a local version of Pinsker's inequality for the KL-divergence between two Bernoulli distributions.

\vspace{1mm}

\begin{lemma}\label{lem:colt}
	There exists a number $K_H > 0$ that only depends on $H$ such that for all $p$ and $n\ge 2$:
	\eqs{
	\PP[ \sum_{i\in p}t_i(n)\kl(\hat\theta_i(n),\theta_i) \geq f_1(n)] \leq K_H n^{-1} (\log(n))^{-2}.
	}
\end{lemma}

\vspace{1mm}
\begin{lemma}[{{\cite[Lemma~2]{garivier2016explore}}}]
\label{lem:refined_Pinsker_Ineq}
For $0\le u<v\le 1$ we have:
\begin{align*}
\kl(u,v)\ge \frac{1}{2v}(u-v)^2.
\end{align*}
\end{lemma}

Next we prove the theorem.\\
\noindent\underline{\textbf{Statement (i):}}
Let $p\in \Pcal$, $n\in \NN$, $t\in \NN^{|E|}$, and $u,\lambda\in (0,1]^{|E|}$ with $u_i\ge \lambda_i$ for all $i$.
By Cauchy-Schwarz inequality we have:
\begin{align*}
p^\top\lambda^{-1} - p^\top u^{-1} &=\sum_{i\in p} \frac{u_i-\lambda_i}{u_i\lambda_i}=\sum_{i\in p} \frac{\sqrt{t_i}(u_i-\lambda_i)}{\sqrt{u_i}} \frac{1}{\lambda_i\sqrt{t_iu_i}} \\
&\le \sqrt{\sum_{i\in p}\frac{t_i(u_i-\lambda_i)^2}{u_i}} \sqrt{\sum_{i\in p}\frac{1}{t_iu_i\lambda_i^2}} \\
&\le \sqrt{\sum_{i\in p}\frac{t_i(u_i-\lambda_i)^2}{u_i}} \sqrt{\sum_{i\in p}\frac{1}{t_i\lambda_i^3}},
\end{align*}
where we used $u_i\ge \lambda_i$ for all $i$ in the last step. Using Lemma~\ref{lem:refined_Pinsker_Ineq}, it then follows that
\begin{align*}
p^\top\lambda^{-1} - p^\top u^{-1} &\le \sqrt{\sum_{i\in p} 2t_i\kl(\lambda_i,u_i)}\sqrt{\sum_{i\in p}\frac{1}{t_i\lambda_i^3}}.
\end{align*}

Thus, $\sum_{i\in p}t_i\kl(\lambda_i,u_i)\leq f_1(n)$ implies:
	\eqs{
		p^\top \lambda^{-1} - p^\top u^{-1} \le \sqrt{\sum_{i\in p} \frac{2f_1(n)}{t_i\lambda_i^3}},
}	
or equivalently, $p^{\top}u^{-1}\ge c_p(n,\lambda,t)$. Hence, by definition of $b_p(n,\lambda,t)$, we have $b_p(n,\lambda,t) \geq c_p(n,\lambda,t)$.	

\vspace{3mm}
\noindent\underline{\textbf{Statement (ii):}}
If $\sum_{i\in p}t_i(n) \kl(\hat\theta_i(n),\theta_i) \leq f_1(n)$, then we have $b_p(n,\hat\theta(n),t(n)) \leq p^\top \theta^{-1}$ by definition of $b_p$. Therefore, using Lemma~\ref{lem:colt}, there exists $K_H$ such that for all $n\ge 2$ we have:
\begin{align*}
\PP[b_p(n,\hat\theta(n),t(n)) &> p^\top \theta^{-1}] \\
&\leq \PP[\sum_{i\in p}t_i(n)\kl(\hat\theta_i(n),\theta_i)\geq f_1(n)] \\
                                 &\leq K_H n^{-1} (\log(n))^{-2},
\end{align*}
which concludes the proof.
\ep

\section{Proof of Theorem \ref{thm:regret_geocombucb}}

\subsection{Preliminary}

Define $a = (1 - 2^{- \frac 14})$ and $\varepsilon = a {\Delta_{\min} \over D^+} < a$. For $s \in \NN^{|E|}$ and $p \in {\cal P}$ define $h(s) = \sum_{i \in p} {1 \over s_i}$. Define $s_i(n)=t_i(n)\hat\theta_i(n)$ the number of packets routed through link $i$ before the $n$-th packet is sent and $ s(n)=(s_i(n))_{i\in E}$. To ease notation define $h(n) = h(s(n))$. We will use the following technical lemma.
\begin{lemma}\label{lem:sumsquare}
Consider $S \subset \NN$, $(s(n))_{n}$ an integer sequence such that $s(n) \ne s(n')$ for all $(n,n') \in S$, $n \ne n'$. Consider a constant $C > 0$, and a positive function $\delta$, such that $\min_{n \in S} \delta(s(n)) \ge \delta_{\min}$. Then:
$$
Z := \sum_{n \in S} \delta(s(n)) \ind \{ s(n) \leq C \delta(s(n))^{-2}  \} \leq {2 C \over \delta_{\min}}.
$$
\end{lemma}
\bp
	If $s(n) \leq C \delta(s(n))^{-2}$, we have  $\delta(s(n)) \le \sqrt{C/s(n)}$, and $s(n) \le C \delta_{\min}^{-2}$. So:
	$$
		Z \le \sum_{n \in S} \sum_{t = 1}^{C \delta_{\min}^{-2}} \ind\{ s(n) = t\} \sqrt{C \over t} \le \sum_{t = 1}^{C \delta_{\min}^{-2}} \sqrt{C \over t},
	$$
	using the fact that $\sum_{n \in S}  \ind\{ s(n) = t\} \le 1$. Using the inequality $\sum_{t=1}^{T} t^{- {1 \over 2}} \le \int_{1}^T t^{- {1 \over 2}} dt \le 2 \sqrt{T}$ yields the result.
\ep

%For any $n\in \mathbb{N}$, {\color{blue}$s\in \mathbb{N}^{|E|}$}, $p\in\Pcal$, and $\lambda\in (0, 1]^{|E|}$ define
%{\color{blue}$$h_{n,s,p,\lambda}=\sqrt{2f_1(n)\sum_{i\in p} \frac{1}{s_i\lambda_i^2}}.$$}
%
%{\color{blue}

% Hence, for all $p$, we have:
%$
%c_p(n)=p^{\top}\hat\theta(n)^{-1}-h_{n,s(n),p,\hat\theta(n)}.
%$
%}
\subsection{Proof of the Theorem}

For any $n$, introduce the following events:
\begin{align*}
A_n&=\Bigl\{\sum_{i\in p^\star} t_i(n)\kl(\hat\theta_i(n), \theta_i)> f_1(n)\Big\},\\%\cup
B_{n,i}&=\{p_i(n)=1,\; |\hat\theta_i(n)-\theta_i|\ge \varepsilon\theta_i\},\;\; B_n=\bigcup_{i\in E} B_{n,i},\\
%C_{n,i}&=\{p_i(n)=1,\; \hat\theta_i(n)\le \theta_i/\delta\},\;\; C_n=\bigcup_{i\in E} C_{n,i},\\
F_n&=\{\Delta_{p(n)} \le (1-a)^{-2} \theta_{\min}^{-1} \sqrt{ 2 f_1(N) h(n)} \}.
\end{align*}
We first prove that $p(n) \ne p^\star$ implies: $n \in A_n \cup B_n \cup F_n$. Consider $n$ such that $p(n) \ne p^\star$ and $A_n \cap B_n$ does not occur. By design of the algorithm, $\xi_{p(n)}(n) \le \xi_{p^\star}(n)$, and $\xi_{p^\star}(n) \le D^\star$ since $A_n$ does not occur. By Theorem \ref{thm:geocombucb_properties} we have $c_{p(n)}(n) \le \xi_{p(n)}(n)$. Hence $c_{p(n)}(n) \le D^\star$. This implies:
$$
p(n)^\top \hat\theta(n)^{-1} - \sqrt{\sum_{i \in p} {2 f_1(n) \over s_i(n) \hat\theta_i(n)^2}} \le D^\star,
$$
so that:
$$
\Delta_{p(n)} \le p(n)^\top \theta^{-1} - p(n)^\top \hat\theta(n)^{-1}  + \sqrt{\sum_{i \in p(n)} {2 f_1(n) \over s_i(n) \hat\theta_i(n)^2}}.
$$
Since $B_n$ does not occur $\hat\theta(n)^{-1} \ge \theta^{-1}/(1 + \varepsilon)$ and:
\als{ p(n)^\top \theta^{-1}- p(n)^\top \hat\theta(n)^{-1}  &\le  {p(n)^\top \theta^{-1} \varepsilon \over (1+\varepsilon) } \le  D^+ \varepsilon \\ &= a \Delta_{\min} \le a \Delta_{p(n)}.
}
Also $\hat\theta_i(n) \ge \theta_{\min} (1 - a)$, and $f_1(n) \le f_1(N)$ so:
$$
 \sum_{i \in p(n)} {2 f_1(n) \over s_i(n) \hat\theta_i(n)^2} \le  {2 f_1(N) h(n) \over (1-a)^2 \theta_{\min}^2}.
$$
Hence:
$$
\Delta_{p(n)} \le a \Delta_{p(n)}  + {\sqrt{ 2 f_1(N) h(n)} \over (1-a) \theta_{\min}},
$$
and $\Delta_{p(n)} \le (1-a)^{-2} \theta_{\min}^{-1} \sqrt{ 2 f_1(N) h(n)}$ and $n \in F_n$.

The regret $R^\pi(N)$ is upper bounded by:
\begin{align*}
\EE\left(\sum_{n=1}^N \Delta_{p(n)}\right) \le \EE\left(\sum_{n=1}^N \Delta_{p(n)} (\ind\{A_n\} + \ind\{B_n\} + \ind\{F_n\})\right).
\end{align*}
\underline{Set $A$}: Using corollary \ref{corr:index_ge_Dstar}, and $K_H \ge 1$ we have:
\eq{\label{setA}
\sum_{n \ge 1} \PP(A_n) \le 1+K_H\sum_{n\ge 2} n^{-1}(\log(n))^{-2} \le 4 K_H.
}
\underline{Set $B$}: Define $\tau_i(n) = \sum_{n'=1}^n \mathbbmss 1\{B_{n',i}\}$. Since $B_{n',i}$ implies $p_i(n')=1$, we have $s_i(n) \geq \tau_i(n)$. Applying \cite[Lemma~B.1]{combes2014unimodal_techreport}, we have $\sum_{n=1}^N \PP(B_{n,i}) \leq 2 (\varepsilon \theta_i)^{-2}$. A union bound yields:
\eq{\label{setB}
	\sum_{n=1}^N \PP(B_{n}) \le 2 \varepsilon^{-2} \sum_{i \in E} \theta_{i}^{-2}.
}
\underline{Set $F$}: Define $U=\frac{4 f_1(N)}{(1-a)^4 \theta_{\min}^2}$. Define the set
$$
S_n =\{i\in p(n) : s_i(n) \le HU \Delta_{p(n)}^{-2} \}
$$
and events:
\als{
G_{n}&=\{|S_n|\ge \sqrt{H} \},\\
L_{n}&=\{|S_n|< \sqrt{H}, \min_{i \in p(n)} s_i(n)\le \sqrt{H} U\Delta_{p(n)}^{-2}]\}.
}
Assume that neither $G_n$ nor $L_n$ occurs, then:
\als{
h(n) &= \sum_{i \in p(n), i \in S_n} {1 \over s_i(n)} + \sum_{i \in p(n), i \notin S_n} {1 \over s_i(n)} \\
&\le {|S_n| \Delta_{p(n)}^2 \over \sqrt{H} U} +{(H - |S_n|) \Delta_{p(n)}^2 \over H U} < {2 \Delta_{p(n)}^2 \over U},
}
since $|S_n| < \sqrt{H}$. Hence $\Delta_{p(n)}^2 > U h(n) / 2$ and $F_n$ does not occur. So $F_n \subset G_n \cup L_n$. Further decompose $G_n$ and $L_n$ as:
\begin{align*}
G_{i,n}&= G_n \cap \{i\in p(n), \;s_i(n) \le HU \Delta_{p(n)}^{-2}\},\\
L_{i,n}&=L_n \cap \{i\in p(n), \;s_i(n) \le  \sqrt{H} U \Delta_{p(n)}^{-2}\}.
\end{align*}
Applying Lemma~\ref{lem:sumsquare} twice, we get:
\als{
	\sum_{n=1}^N \Delta_{p(n)} \ind\{G_{i,n}\} \leq { HU \over \Delta_{\min}}  \text{    ,    }
	\sum_{n=1}^N \Delta_{p(n)} \ind\{L_{i,n}\} \leq { \sqrt{H} U \over \Delta_{\min}}.
}
We have
\eqs{
\sum_{i\in E} \mathbbmss 1 \{G_{i,n}\} = |S_n|\mathbbmss 1 \{G_n\}\ge \sqrt{H} \mathbbmss 1 \{G_n\}.
}
So:
\eqs{
\sum_{n=1}^N \Delta_{p(n)} \ind\{G_{n}\} \le {1 \over \sqrt{H}} \sum_{n=1}^N \sum_{i \in E} \Delta_{p(n)} \ind\{G_{i,n}\} \le {|E| \sqrt{H} U \over \Delta_{\min}}.
}
Further:
\eqs{
	\sum_{n=1}^N \Delta_{p(n)} \ind\{L_{n}\} \le \sum_{n=1}^N \sum_{i \in E} \Delta_{p(n)} \ind\{L_{i,n}\} \le { |E| \sqrt{H} U \over \Delta_{\min}}.
}
Since $\ind\{F_n\} \le \ind\{G_n\} + \ind\{L_n\}$ we get:
\eq{\label{setF}
\EE\left(\sum_{n=1}^N \Delta_{p(n)} \ind\{F_n\} \right) \le {2 |E| \sqrt{H}U \over \Delta_{\min}}.
}

Combining \eqref{setA}, \eqref{setB} and \eqref{setF} with $\Delta_{p(n)}\le D^+$, yields the announced result:
$$
R^\pi(N) \le {2 |E| \sqrt{H}U \over \Delta_{\min}} + D^+ \left( 4 K_H + 2 \varepsilon^{-2} \sum_{i \in E} \theta_{i}^{-2} \right).
$$
\ep
\section{Proof of Theorem \ref{thm:regret_geocombucb_2}}

The proof technique is similar to the analysis of~\cite[Theorem~5]{kveton2014tight}.
\subsection{Preliminary}
For $s \in \NN^{|E|}$ and $p \in {\cal P}$ define $h'(s) = (\sum_{i \in p} {1 \over \sqrt{s_i}})^2$, and as before $s_i(n)=t_i(n)\hat\theta_i(n)$ and $s(n)=(s_i(n))_{i\in E}$, and $h'(n) = h'(s(n))$.
We will use the following technical lemma.
\begin{lemma} \label{lem:index_KL_SR} For all $n,t\in \NN$, $\lambda\in(0,1]$, and $i\in E$:
\als{\omega_i(n,\lambda,t)\ge \frac{1}{\lambda}-\sqrt{\frac{2f_2(n)}{t\lambda^3}}.}
\end{lemma}
\bp
Let $i\in E$, $n,t\in \NN$ and $u,\lambda\in (0,1]$ with $u\ge \lambda$.
We have:
\begin{align*}
\frac{1}{\lambda}-\frac{1}{u}=\sqrt{\frac{t(u-\lambda)^2}{u}}\cdot\frac{1}{\sqrt{tu\lambda^2}}
\le \sqrt{2t\kl(\lambda,u)}\cdot\frac{1}{\sqrt{t\lambda^3}},
\end{align*}
where the second inequality follows from Lemma~\ref{lem:refined_Pinsker_Ineq} and $u\ge \lambda$. Hence, $t\kl(\lambda,u) \leq f_2(n)$ implies:
$
\frac{1}{u}\ge \frac{1}{\lambda}-\sqrt{\frac{2f_2(n)}{t\lambda^3}}.
$
The above holds for all $u\in [\lambda,1]$, and by definition of $\omega_i(n,\lambda,t)$:
$$
\omega_i(n,\lambda,t)\ge \frac{1}{\lambda}-\sqrt{\frac{2f_2(n)}{t\lambda^3}}.
$$
\ep

\subsection{Proof of the theorem}
For any $n$, we define the following events:
\als{
A_{n,i}&= \Bigl\{ t_i(n)\kl(\hat\theta_i(n), \theta_i)> f_2(n)\Big\},\;\; A_n=\bigcup_{i\in p^\star} A_{n,i},\\
B_{n,i}&=\{p_i(n)=1,\; |\hat\theta_i(n)-\theta_i|\ge \varepsilon\theta_i\},\;\; B_n=\bigcup_{i\in E} B_{n,i},\\
F_n&=\{\Delta_{p(n)} \le (1-a)^{-2} \theta_{\min}^{-1} \sqrt{ 2 f_2(N) h'(n)} \}.
}

We show that $p(n) \ne p^\star$ implies: $n \in A_n \cup B_n \cup F_n$. Consider $n$ such that $p(n) \ne p^\star$ and $A_n \cup B_n$ does not occur. By design of the algorithm, $p(n)^\top \omega(n) \le (p^\star)^\top \omega(n)$, and $(p^\star)^\top \omega(n) \le D^\star$ since $A_n$ does not occur. Hence $p(n)^\top \omega(n) \le D^\star$. By Lemma~\ref{lem:index_KL_SR}, for all $i$:
\eqs{
\omega_i(n)\ge \frac{1}{\hat\theta_i(n)}-\sqrt{\frac{2f_2(n)}{s_i(n)\hat\theta_i(n)^2}}.
}
Summing over $i \in p(n)$ we get:
$$
\Delta_{p(n)} \le  p(n)^\top \theta^{-1} - p(n)^\top \hat\theta(n)^{-1}  + \sum_{i \in p(n)} \sqrt{2 f_2(n) \over s_i(n) \hat\theta_i(n)^2}.
$$
As before, when $B_n$ does not occur we have
$$
p(n)^\top \theta^{-1} - p(n)^\top \hat\theta(n)^{-1} \le a \Delta_{p(n)}.
$$
Furthermore $\hat\theta_i(n) \ge \theta_{\min}(1-a)$ and $f_2(n) \le f_2(N)$ so that:
$$
 \sum_{i \in p(n)} \sqrt{2 f_2(n) \over s_i(n) \hat\theta_i(n)^2} \le  \sum_{i \in p(n)} \sqrt{f_2(N) \over s_i(n) \theta_{\min}^2(1 - a)^2 },
$$
Hence:
$$
\Delta_{p(n)} \le a \Delta_{p(n)}   +  {\sqrt{ 2 f_2(N) h'(n)} \over (1-a) \theta_{\min}}
$$
and $\Delta_{p(n)} \le (1-a)^{-2} \theta_{\min}^{-1} \sqrt{ 2 f_2(N) h'(n)}$ so that $n \in F_n$.

The regret $R^\pi(N)$ is upper bounded by:
\begin{align*}
\EE\Big(\sum_{n=1}^N \Delta_{p(n)}\Big) \le \EE\Big(\sum_{n=1}^N \Delta_{p(n)} (\ind\{A_n\} + \ind\{B_n\} + \ind\{F_n\})\Big).
\end{align*}

\underline{Set $A$}: By \cite[Theorem~10]{Garivier2011} and a union bound:
\eqs{
\PP(A_{n}) \le \sum_{i \in p^\star} \PP(A_{n,i}) \le H \lceil f_2(n)\log(n)\rceil e^{1-f_2(n)}.
}
Hence:
\eq{\label{setA2}
\sum_{n=1}^N \PP(A_{n}) \le H \Big(1+e\sum_{n\ge 2} \lceil f_2(n)\log(n)\rceil e^{-f_2(n)} \Big) \le 8|H|.
}

\underline{Set $B$}: As in the proof of Theorem~\ref{thm:regret_geocombucb}:
\eq{\label{setB2}
	\sum_{n=1}^N \PP(B_{n}) \le 2 \varepsilon^{-2} \sum_{i \in E} \theta_{i}^{-2}.
}

\underline{Set $F$}: Define $U'= 2 H^2 f_2(N) (1-a)^{-4} \theta_{\min}^{-2}$. Similarly to the proof of \cite[Theorem~5]{kveton2014tight}, consider $\alpha,\beta > 0$, for $\ell \in \NN$ define $\alpha_\ell=\left(\frac{1-\beta}{\sqrt{\alpha}-\beta}\right)^2 \alpha^\ell$ and $\beta_\ell=\beta^\ell$. Introduce set $S_{\ell,n}$ and events $G_{\ell,n}$:
\als{
S_{\ell,n} &= \{i \in p(n) , s_i(n) \le  U' \alpha_{\ell} \Delta_{p(n)}^{-2}\},\\
G_{\ell,n} &=  \{|S_{\ell,n}|\ge \beta_{\ell}H\} \cap \{|S_{j,n}|<\beta_j H , j=1,...,\ell-1 \}.
}

If $\overline{ \cup_{\ell \ge 1} G_{\ell,n} } = \{|S_{\ell,n}|< H\beta_\ell , \ell \ge 1 \}$ then:
\als{
	 \sum_{\ell \ge 1} {|S_{\ell-1,n}| - |S_{\ell,n}| \over \sqrt{\alpha_{\ell}}}
	&= {|S_{0,n}| \over \sqrt{\alpha_{1}}} + \sum_{\ell \ge 1} |S_{\ell,n}|\Big( {1 \over \sqrt{\alpha_{\ell+1}}} - {1 \over \sqrt{\alpha_{\ell}}} \Big) \\
	&< 	{H \beta_0 \over \sqrt{\alpha_{1}}} + \sum_{\ell \ge 1} H \beta_{\ell} \Big({1 \over \sqrt{\alpha_{\ell+1}}} - {1 \over \sqrt{\alpha_{\ell}}} \Big)  \sk
	&=  H \sum_{\ell \ge 1}  {\beta_{\ell} - \beta_{\ell - 1} \over \sqrt{ \alpha_{\ell}}} \le  H,
}
since ${1 \over \sqrt{\alpha_{\ell+1}}} - {1 \over \sqrt{\alpha_{\ell}}} \ge 0$. Now:
\als{
|\{i:  s_i(n) \in U' \Delta_{p(n)}^{-2} [\alpha_\ell,\alpha_{\ell-1}] \}| = |S_{\ell-1,n}| - |S_{\ell,n}|
}
so that:
\eqs{
\sqrt{h'(n)} \le \sum_{\ell \ge 1} {(|S_{\ell-1,n}| - |S_{\ell,n}|) \over  \sqrt{\alpha_{\ell}}} {\Delta_{p(n)} \over \sqrt{U'}} < H {\Delta_{p(n)} \over \sqrt{U'}}.
}
Hence $\Delta_{p(n)}^2 >  h'(n) U' H^{-2}$, and $F_n$ does not occur. Therefore $F_n \subset \cup_{\ell \ge 1} G_{\ell,n}$ and:
\eqs{
\sum_{n=1}^N  \Delta_{p(n)}  \ind \{F_n\} \le \sum_{n=1}^N \sum_{\ell \ge 1} \Delta_{p(n)}  \ind \{G_{\ell,n} \}.
}
Further decompose $G_{i,\ell}$ as:
$$
G_{i,\ell,n}=G_{\ell,n}\cap \{i\in p(n), \; s_i(n)\le U' \alpha_{\ell} \Delta_{p(n)}^{-2}\}.
$$
Observe that:
\eqs{
 \bI\{G_{\ell,n}\} \le {|S_{\ell,n}| \over H \beta_{\ell} }\bI\{G_{\ell,n}\} = {1 \over H \beta_{\ell} } \sum_{i\in E} \ind \{G_{i,\ell,n} \}.
}
Applying Lemma~\ref{lem:sumsquare}, we get:
\als{
	\sum_{n=1}^N \Delta_{p(n)} \ind \{G_{i,\ell,n} \} &\le \sum_{n=1}^N \Delta_{p(n)} \ind \left\{s_i(n)\le {U' \alpha_{\ell}  \over \Delta_{p(n)}^{2}} \right\} \sk
	&\le {2U' \alpha_\ell \over \Delta_{\min}}.
}
Putting it together:
\eq{\label{setF2}
\sum_{n=1}^N  \Delta_{p(n)}  \ind \{ F_n \} \le {2 |E| U' \over H \Delta_{\min}} \sum_{\ell \ge 1} { \alpha_{\ell} \over  \beta_{\ell}} \le {90 |E| U' \over H \Delta_{\min}},
}
by choosing $\alpha = 0.15$ and  $\beta = 0.24$ so that $\sum_{\ell \ge 1} {\alpha_\ell \over \beta_{\ell}} \le 45$.

Combining \eqref{setA2}, \eqref{setB2} and \eqref{setF2} with $\Delta_{p(n)}\le D^{+}$, yields the result:
$$
R^\pi(N) \le {90 |E| U' \over H \Delta_{\min}} + D^{+} \left( 8 H + 2 \varepsilon^{-2} \sum_{i \in E} \theta_{i}^{-2} \right).
$$
\ep

\section{Proof of Proposition~\ref{prop:GeoCombUCB_Line}}
\label{sec:AlgLineProof}

%In the line network, the shortest-path routing problem is simplified since the decisions at different hops are decoupled.It suffices to route packets on the best link on each hop, and the total regret is the summation of the regrets on all hops.
In the line network, \textsc{KL-SR} simply chooses the link with the smallest index on each hop. Hence, on each hop,
\textsc{KL-SR} is equivalent to the KL-UCB algorithm for a classical MAB with geometrically distributed rewards.  By~\cite[Theorem~1~and~Lemma~6]{Garivier2011}, the regret of KL-SR on the $m$-th hop asymptotically grows as:
\begin{align*}
\sum_{i \in E_m\setminus p^\star}
\frac{\log(N)}{\klg(\theta_i, \theta_{\zeta(i)})}\left(\frac{1}{\theta_i} -\frac{1}{\theta_{\zeta(i)}} \right),
\end{align*}
where $E_m$ denotes the set of links in the $m$-th hop.
Since decisions at various hops are decoupled, the regret due to all hops satisfies
\begin{align*}
\limsup_{N \rightarrow \infty} \dfrac{R^{\text{KL-SR}}(N)}{\log(N)}
&\leq  \sum_{m=1}^H \sum_{i\in E_m\setminus p^\star}
\dfrac{ \frac{1}{\theta_i} - \frac{1}{\theta_{\zeta(i)}} }
{ \klg(\theta_i, \theta_{\zeta(i)}) } \sk
&= \sum_{i\notin p^\star}
\dfrac{ \frac{1}{\theta_i} - \frac{1}{\theta_{\zeta(i)}} }
{ \klg(\theta_i, \theta_{\zeta(i)}) }=c_2(\theta).
\end{align*}

Furthermore, using Lemma~\ref{lem:KLG_KL} and Lemma~\ref{lem:refined_Pinsker_Ineq} we have for any $i\notin p^\star$:
\begin{align*}
\dfrac{ {\frac{1}{\theta_i}}  - \frac{1}{\theta_{\zeta(i)}} }
{  \klg(\theta_i, \theta_{\zeta(i)}) }&= \frac{\theta_{\zeta(i)}-\theta_i}{\theta_{\zeta(i)} \kl(\theta_i,\theta_{\zeta(i)})} \le \frac{2}{\theta_{\zeta(i)}-\theta_i}.
% \\&= \frac{2}{\theta_{\zeta(i)} \theta_i}\cdot \frac{1}{\theta_i^{-1}-\theta_{\zeta(i)}^{-1}}.
\end{align*}
Moreover, in line networks $\Delta_{\min}=\min_{i\notin p^\star} (\theta_i^{-1}-\theta_{\zeta(i)}^{-1})$. Thus,
\begin{align*}
c_2(\theta)&\le \sum_{i\notin p^\star} \frac{2}{\theta_{\zeta(i)}-\theta_i} = \sum_{i\notin p^\star} \frac{2}{\theta_i\theta_{\zeta(i)}(\theta_i^{-1}-\theta_{\zeta(i)}^{-1})} \\
&\le \frac{|E|-H}{\Delta_{\min}}\cdot\frac{2}{\min_{i\notin p^\star} \theta_i\theta_{\zeta(i)}} \le \frac{2(|E|-H)}{\Delta_{\min}\theta_{\min}^2},
\end{align*}
which completes the proof.

\ep

\section{Proof of Proposition \ref{prop:index_computation}} \label{sec:index_computation}
%{\color{red} It is easy to verify that the function $F(\lambda)$ is strictly increasing.}
%The rest of the proof follows the similar lines as in the proof of \cite[Theorem~4]{combes2015stochastic}.

The proof is similar to that of \cite[Theorem~4]{combes2015stochastic}.
Note that if $i\notin I_p(\lambda)$, then the optimal solution satisfies $u_i=1$ since $\kl(1, v) = \infty$ unless $v = 1$.
Thus, if $I_p(\lambda)=\emptyset$, then $u_i=1, \forall i\in E$, and $b_p(n,\lambda,t)=\sum_{i\in p} p_i$.

If $I_p(\lambda) \neq \emptyset$, let $i\in  I_p(\lambda)$. Computing $b_p$ involves solving a convex optimization problem with one inequality constraint which must hold with equality since $u_i \mapsto \kl(\lambda_i, u_i)$ is monotone increasing for $u_i \geq \lambda_i$. Since $\frac{d}{du_i}\kl(\lambda_i,u_i) = {u - \lambda \over u(1-u)}$, the Karush-Kuhn-Tucker conditions are:
\begin{align*}
&\frac{1}{{u_i}^2} - \gamma t_i {u_i - \lambda_i \over u_i(1-u_i)}=0,\\ %\quad
&\sum_{i \in I_p(\lambda)} t_i \kl( \lambda_i, u_i)  - f_1(n) = 0.
\end{align*}
with $\gamma>0$ the Lagrange multiplier. The first equation is the quadratic equation:
\eqs{
u_i^2 + u_i\left(\frac{1}{\gamma t_i}-\lambda_i \right) - \frac{1}{\gamma t_i}=0.
}
Solving for $u_i$, we obtain $u_i(\gamma) = g(\gamma,\lambda_i,t_i)$ and replacing in the second equation, we obtain $F(\gamma,n,\lambda,t) = f_1(n)$. The results then follow directly.
\ep

\section{Regret Upper Bound for CUCB} \label{sec:CUCB_regret}

CUCB (see \cite{chen2013combinatorial_icml}) uses the following link index:
$$
\gamma_i(n)=\frac{1}{\hat\theta_i(n)+\sqrt{1.5\log(n)/t_i(n)}} \ \ , \ \ \forall i \in E
$$
Define $a = (1 - 2^{- \frac 14})$ and $\varepsilon = a {\Delta_{\min} \over D^+} < a$. For any $s\in \mathbb{N}^{|E|}$ and $p\in \Pcal$ define $h'(s)=(\sum_{i\in p} \frac{1}{\sqrt{s_i}})^2$, and as in the proof of Theorem 5.4, $s_i(n)=t_i(n)\hat\theta_i(n)$ and $s(n)=(s_i(n))_{i\in E}$, and $h'(n)=h'(s(n))$. We have that:
\begin{align}
&p(n)^\top \gamma(n)=\sum_{i\in p(n)}\frac{1}{\hat\theta_i(n)+\sqrt{1.5\hat\theta_i(n)\log(n)/s_i(n)}} \sk
&=\sum_{i\in p(n)}\frac{1}{\hat\theta_i(n)}-\sum_{i\in p(n)}\frac{\sqrt{1.5\log(n)/(s_i(n)\hat\theta_i(n)^3)}}{1+\hat\theta_i(n)^{-\frac{1}{2}} \sqrt{1.5\log(n)/s_i(n)} } \sk
\label{eq:CUCBindex_LB}
&\ge p(n)^\top\hat\theta(n)^{-1}-\sum_{i\in p(n)} \sqrt{\frac{1.5\log(n)}{s_i(n)\hat\theta_i(n)^3}}.
\end{align}

For any $n$, introduce the following events:
\begin{align*}
A_{n,i}&=\Bigl\{|\hat\theta_i(n)-\theta_i|> \sqrt{1.5\log(n)/t_i(n)}\Big\},\;\; A_n=\bigcup_{i\in p^\star} A_{n,i},\\
B_{n,i}&=\{p_i(n)=1,\; |\hat\theta_i(n)-\theta_i|\ge \varepsilon\theta_i\},\;\; B_n=\bigcup_{i\in E} B_{n,i},\\
F_n&=\{\Delta_{p(n)} \le (1-a)^{-\frac{5}{2}} \theta_{\min}^{-\frac{3}{2}} \sqrt{ 2 \log(N) h'(n)} \}.
\end{align*}

We show that if $p(n) \ne p^\star$ then $A_n \cup B_n \cup F_n$ occurs. Consider $n$ such that $p(n) \ne p^\star$ and $A_n \cup B_n$ does not occur. By design of the algorithm, $p(n)^\top \gamma(n) \le (p^\star)^\top \gamma(n)$, and $(p^\star)^\top \gamma(n) \le D^\star$ since $A_n$ does not occur. Hence $p(n)^\top \gamma(n) \le D^\star$.

When $B_n$ does not occur, $(1-a)\theta_{\min}\le \hat\theta_i(n)\le (1+\varepsilon)\theta_i$ and $p(n)^\top\theta^{-1}-p(n)^\top\hat\theta(n)^{-1}\le a\Delta_{p(n)}$. Hence, using (\ref{eq:CUCBindex_LB}), we get
\begin{align*}
\Delta_{p(n)}&=p(n)^\top\theta^{-1}-D^\star\le p(n)^\top\theta^{-1}-p(n)^\top\gamma(n) \sk
&\le a\Delta_{p(n)} + (1-a)^{-\frac{3}{2}}\theta_{\min}^{-\frac{3}{2}}\sqrt{1.5\log(N)h'(n)} \sk
\end{align*}
%\begin{align*}
%&p(n)^\top \gamma(n)\ge (1+\varepsilon)^{-1}p(n)^\top\theta^{-1}-\sum_{i\in p(n)} \sqrt{\frac{1.5\log(n)}{s_i(n)(1-a)^3\theta_{\min}^3}} \sk
%&\ge (1+\varepsilon)^{-1}p(n)^\top\theta^{-1} - (1-a)^{-1.5}\theta_{\min}^{-1.5}\sqrt{1.5\log(N)h'(n)}
%\end{align*}
%
%Hence, when $A_n\cap B_n$ does not occur, $(p^\star)^\top\gamma(n)\le D^\star$ and hence,
%$$
%p(n)^\top\hat\theta(n)^{-1}-\sum_{i\in p(n)} \sqrt{\frac{1.5\log(n)}{s_i(n)\hat\theta_i(n)^3}} \le D^\star.
%$$
%
%
%
%Hence,
so that
$$
\Delta_{p(n)} \le (1-a)^{-\frac{5}{2}} \theta_{\min}^{-\frac{3}{2}}\sqrt{1.5\log(N)h'(n)}
$$
and thus $n\in F_n$.

% Note that $\varepsilon D^\star \le a\frac{\Delta_{\min}}{D^+} D^\star \le a\Delta_{\min}\le a\Delta_{p(n)}$. Hence

%As before, when $B_n$ occurs we have
%$$
%p(n)^\top \theta^{-1} - p(n)^\top \hat\theta(n)^{-1} \le a \Delta_{p(n)}.
%$$
%Furthermore $\hat\theta_i(n) \ge \theta_{\min}(1-a)$ so that:
%$$
% \sum_{i \in p(n)} \sqrt{2 \log(n) \over s_i(n) \hat\theta_i(n)^2} \le  \sum_{i \in p(n)} \sqrt{\log(N) \over s_i(n) \theta_{\min}^2(1 - a)^2 },
%$$
%Hence:
%$$
%\Delta_{p(n)} \le a \Delta_{p(n)}   +  {\sqrt{ 2 \log(N) h'(n)} \over (1-a) \theta_{\min}}
%$$
%and

The regret $R^\pi(N)$ is upper bounded by:
\begin{align*}
\EE\Big(\sum_{n=1}^N \Delta_{p(n)}\Big) \le \EE\Big(\sum_{n=1}^N \Delta_{p(n)} (\ind\{A_n\} + \ind\{B_n\} + \ind\{F_n\})\Big).
\end{align*}

\underline{Set $A$}:
Using a Chernoff bound and a union bound, we have that $\PP(A_n)\le 2Hn^{-2}$ (see, e.g., \cite[Lemma~3]{chen2013combinatorial_icml}). Hence
\begin{align}
\label{setA3}
\sum_{n=1}^N \PP(A_n)\le \sum_{n=1}^N \frac{2H}{n^2}\le \frac{2\pi^2H}{3}.
\end{align}

\underline{Set $B$}:
As in the proof of Theorem~\ref{thm:regret_geocombucb}:
\eq{\label{setB3}
	\sum_{n=1}^N \PP(B_{n}) \le 2 \varepsilon^{-2} \sum_{i \in E} \theta_{i}^{-2}.
}

\underline{Set $F$}:
Define $U'= 2 H^2 f_2(N) (1-a)^{-\frac{5}{2}} \theta_{\min}^{-3}$. By the same technique as the proof of Theorem~\ref{thm:regret_geocombucb_2} we get
\begin{align}
\label{setF3}
\sum_{n=1}^N \Delta_{p(n)}\mathbbmss 1\{F_n\}\le \frac{278H|E|\log(N)}{\Delta_{\min}\theta_{\min}^3}.
\end{align}
Putting  (\ref{setA3}), (\ref{setB3}), and  (\ref{setF3}) together, we obtain
\begin{align*}
R^\pi(N)\le  \frac{278H|E|\log(N)}{\Delta_{\min}\theta_{\min}^3}
+ 2 D^+\left(\frac{\pi^2 H}{3}+ \sum_{i\in E}{1 \over (\varepsilon\theta_{i})^{2}}\right).
\end{align*}
\ep

\begin{IEEEbiography}[{\includegraphics[width=1in,height=1.25in,clip,keepaspectratio]{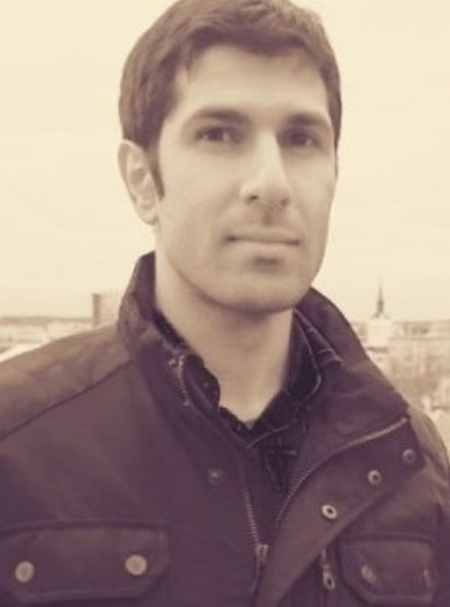}}]{M. Sadegh Talebi}
received his B.S.~in electrical engineering from Iran University of Science and Technology (IUST), Iran, in 2004, his M.Sc.~degree in electrical engineering from Sharif University of Technology, Iran, in 2006.
He is currently pursuing his Ph.D.~in the Department of Automatic Control at KTH The Royal Institute of Technology, Sweden.
%Before starting his Ph.D., he served as a research engineer in School of Computer Science, Institute for Research in Fundamental Sciences (IPM), Iran, where he worked on performance evaluation of networked systems.
His current research interests include resource allocation in networks, sequential decision making, and learning theory.
\end{IEEEbiography}

\begin{IEEEbiography}[{\includegraphics[width=1in,height=1.25in,clip,keepaspectratio]{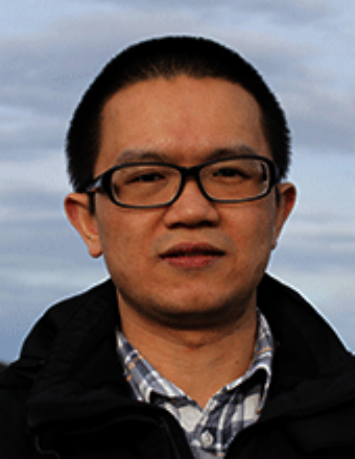}}]{Zhenhua Zou}
received the M.E.~degree from Southeast University, Nanjing, China, and the M.Sc.~degree in communication engineering \textit{(summa cum laude)} from Politecnico di Torino, Torino, Italy, in March 2009 and September 2009, respectively. He received his Ph.D.~degree in telecommunications at the School of Electrical Engineering, KTH The Royal Institute of Technology, Stockholm, Sweden in 2014.
He is now a system engineer at Qamcom Research and Technology in Sweden. His research interest includes algorithm development for real-time communication in wireless lossy networks.
\end{IEEEbiography}

\begin{IEEEbiography}[{\includegraphics[width=1in,height=1.25in,clip,keepaspectratio]{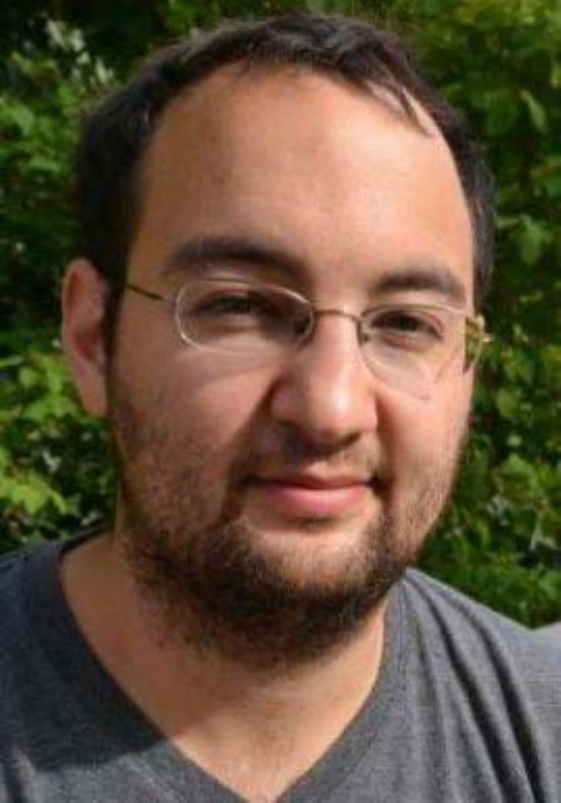}}]{Richard Combes}
%is currently an assistant professor in Supelec. He received the Engineering Degree from Telecom Paristech (2008), the Master Degree in Mathematics from university of Paris VII (2009), and the Ph.D.~degree in Mathematics from university of Paris VI (2013).
% He was a visiting scientist at INRIA (2012) and a post-doc in KTH (2013). He received the best paper award at CNSM 2011. His current research interests are machine learning, networks, and probability.
is  currently  an  Assistant  Professor  with Supelec, Gif-sur-Yvette Cedex, France.
received  the  B.E.~degree  from T\'{e}l\'{e}com   ParisTech,   Paris,   France,   in   2008;   the
Master’s degree in mathematics from Paris Diderot University -- Paris  7,  Paris,  in  2009;  and  the  Ph.D.~degree  in  mathematics  from  the  Pierre-and-Marie-Curie University, Paris, in 2012.
He was a Visiting Scientist  with  the  French  Institute  for  Research  in
Computer Science and Automation (INRIA) in 2012 and a Postdoctoral Researcher with KTH The Royal Institute of Technology, Stockholm, Sweden, in
2013.  He received the Best Paper Award at the 2011 International Conference on Network and Service Management. His
current research interests are machine learning, networks, and probability.
\end{IEEEbiography}

\begin{IEEEbiography}[{\includegraphics[width=1in,height=1.25in,clip,keepaspectratio]{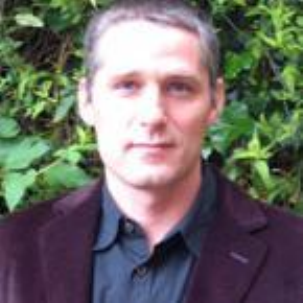}}]{Alexandre Proutiere} received the degree in mathematics from \'{E}cole Normale Sup\'{e}rieure, Paris, France; the degree in engineering from T\'{e}l\'{e}com  ParisTech, Paris, France; and the Ph.D.~degree in applied mathematics from \'{E}cole Polytechnique, Palaiseau, France, in 2003.
He is an Engineer from Corps of Mines. In 2000,
he joined France Telecom R\&D as a Research Engineer. From 2007 to 2011, he was a Researcher at
Microsoft Research, Cambridge, U.K. He is currently a Professor in the Department of Automatic Control at KTH The Royal Institute of Technology, Stockholm, Sweden.
He was the recipient in 2009 of the ACM Sigmetrics Rising Star Award, and received the Best Paper Awards at ACM Sigmetrics conference in 2004 and 2010, and at the ACM Mobihoc Conference in 2009. He was an Associate Editor of \textsc{IEEE/ACM Transactions on Networking} and an editor of \textsc{IEEE Transactions on Control of Network Systems}, and is currently an editor of \textit{Queuing Systems}.
\end{IEEEbiography}

\begin{IEEEbiography}[{\includegraphics[width=1in,height=1.25in,clip,keepaspectratio]{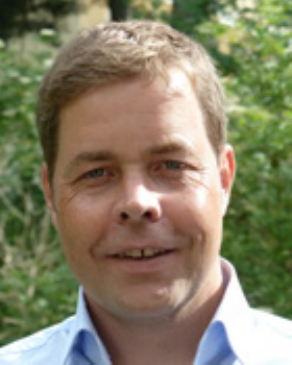}}]{Mikael Johansson}
received the M.Sc.~and Ph.D.~degrees in electrical engineering from Lund University,
Lund, Sweden, in 1994 and 1999, respectively.
He held postdoctoral positions at Stanford University, Stanford, CA, USA, and University of California, Berkeley, CA, USA, before joining KTH The Royal Institute of Technology, Stockholm, Sweden in 2002, where he now serves as Full Professor. He has published two books and more than a hundred
papers, several which are highly cited and have received recognition in terms of best paper awards. He has served on the editorial boards of \textit{Automatica} and the \textsc{IEEE Transactions on Control of Network Systems}, as well as on the program committee
for several top-conferences organized by IEEE and ACM. He has played a leading role in several national and international research projects in control and communications.
\end{IEEEbiography}

\end{document}